\documentclass[twocolumn]{aastex631}

\makeatletter

\makeatletter

\begin{document}

\title{Galaxy cluster characterization with machine learning techniques}

\author{M. Sadikov}
\affiliation{D\'{e}partement de physique, Universit\'{e} de Montr\'{e}al, \\
C.P. 6128 Succ. Centre-ville, Montr\'{e}al \\
H3C 3J7, Canada}
\affiliation{Centre de recherche en astrophysique du Québec (CRAQ)}

\author{J. Hlavacek-Larrondo}
\affiliation{D\'{e}partement de physique, Universit\'{e} de Montr\'{e}al, \\
C.P. 6128 Succ. Centre-ville, Montr\'{e}al \\
H3C 3J7, Canada}
\affiliation{Centre de recherche en astrophysique du Québec (CRAQ)}

\author{L. Perreault-Levasseur}
\affiliation{D\'{e}partement de physique, Universit\'{e} de Montr\'{e}al, \\
C.P. 6128 Succ. Centre-ville, Montr\'{e}al \\
H3C 3J7, Canada}
\affiliation{Centre de recherche en astrophysique du Québec (CRAQ)}
\affiliation{Mila - Québec Artificial Intelligence Institute, Montr\'{e}al, Canada}
\affiliation{Center for Computational Astrophysics, Flatiron Institute,\\
162 5th Avenue, New York, NY 10010, USA}

\author{C. L. Rhea}
\affiliation{D\'{e}partement de physique, Universit\'{e} de Montr\'{e}al, \\
C.P. 6128 Succ. Centre-ville, Montr\'{e}al \\
H3C 3J7, Canada}
\affiliation{Centre de recherche en astrophysique du Québec (CRAQ)}

\author{M. McDonald}
\affiliation{Kavli Institute for Astrophysics and Space Research,\\
Massachusetts Institute of Technology, Cambridge, MA 02139, USA}

\author{M. Ntampaka}
\affiliation{Space Telescope Science Institute, Baltimore, MD 21218, USA}
\affiliation{Department of Physics \& Astronomy, Johns Hopkins University, Baltimore, MD 21218, USA}

\author{J. ZuHone}
\affiliation{Center for Astrophysics | Harvard \& Smithsonian, Cambridge, MA 02138 USA}



\begin{abstract}

We present an analysis of the X-ray properties of the galaxy cluster population in the $z=0$ snapshot of the IllustrisTNG simulations, utilizing machine learning techniques to perform clustering and regression tasks. We examine five properties of the hot gas (the central cooling time, the central electron density, the central entropy excess, the concentration parameter, and the cuspiness) which are commonly used as classification metrics to identify cool core (CC), weak cool core (WCC) and non cool core (NCC) clusters of galaxies. Using mock \textit{Chandra} X-ray images as inputs, we first explore an unsupervised clustering scheme to see how the resulting groups correlate with the CC/WCC/NCC classification based on the different criteria. We observe that the groups replicate almost exactly the separation of the galaxy cluster images when classifying them based on the concentration parameter. We then move on to a regression task, utilizing a ResNet model to predict the value of all five properties.   The network is able to achieve a mean percentage error of 1.8\% for the central cooling time, and a balanced accuracy of 0.83 on the concentration parameter, making them the best-performing metrics. Finally, we use simulation-based inference (SBI) to extract posterior distributions for the network predictions. Our neural network simultaneously predicts all five classification metrics using only mock Chandra X-ray images. 

This study demonstrates that machine learning is a viable approach for analyzing and classifying the large galaxy cluster datasets that will soon become available through current and upcoming X-ray surveys, such as eROSITA.

\end{abstract}



\section{Introduction\label{sec:intro}}
\vspace{3mm}
Galaxy clusters are the largest virialized structures of our universe. They are made up of hundreds to thousands of galaxies, dark matter, and intracluster gas known as the intracluster medium (ICM). The dark matter makes up $\sim$ 84\% of the mass, the galaxies make up $\sim$ 3\% of the mass, and the ICM makes up $\sim$ 13\% of the mass (\citealt{mohr_1999_icm}, \citealt{Vikhlinin_2006_icm}, \citealt{Umetsu_2009_icm}).\\
\\
X-ray observations of galaxy clusters probe the hot ICM. The ICM is a diffuse gas, mainly made up of ionized hydrogen and helium, but also containing iron, silicon, sulfur, argon, carbon and nickel (e.g. \citealt{Loewenstein_2003_chemical_comp}). Due to its high temperature ($ \approx 10^7$ K), the ICM emits strongly at X-ray wavelengths through processes such as Bremsstrahlung emission. Galaxy clusters are often separated into three classes based on the distribution of the ICM within the cluster (e.g. \citealt{million2009chandra}, \citealt{Hudson_2010_article}): cool core clusters (CC),  weak cool core clusters (WCC), and non-cool core clusters (NCC). In cool core clusters, the ICM is strongly concentrated towards the center of the cluster (i.e. the inner $\sim$100 kpc;  \citealt{White_1997}, \citealt{Hudson_2010_article}, \citealt{McDonald_2013_article}), resulting in strong X-ray emission. This leads to a temperature decrease in the core, which results in a pressure decrease as well. However, the gas in the center of the cluster must be able to support the weight of the outer layers, and therefore it becomes denser to prevent the outer layers from collapsing, restoring the pressure support against gravity. This creates a cooling flow, where cool gas is infalling towards the center of the cluster. This cool gas should lead to a star formation rate on the order of 100-1000 $M_{\odot}$/yr (e.g. \citealt{Peterson_Fabian_2006}). However, the observed star formation rate of $\sim$ 10-300 $M_{\odot}$/yr (e.g. \citealt{McNamara_2004}, \citealt{mcdonald_2015_sfr})  suggests that some mechanism is offsetting the cooling. That mechanism is most likely active galactic nucleus (AGN) feedback from the brightest cluster galaxy (BCG); (see review by \citealt{AGN_review_JHL_Li_Churazov}). CC clusters therefore give an important insight into the way the supermassive back hole (SMBH) at the center of the BCG interacts with the surrounding gas through AGN feedback processes. On the other hand, clusters without a strongly peaked core density are non-cool core clusters. NCCs are also more disturbed, while CCs are generally relaxed. Finally, we often add a third category of weak cool core clusters as an intermediate between the two (e.g.  \citealt{Bauer_2005}, \citealt{Sun_2009}, \citealt{Hudson_2010_article}, \citealt{Bharadwaj_2014}).\\
\\
Building on this foundation, recent efforts have turned to leveraging cosmological hydrodynamical simulations, such as the IllustrisTNG, to further dissect the properties of galaxy clusters. This transition from observational studies to simulation-based analyses is driven by the need to complement real-world observations with the predictive power of simulations, which can explore a vast parameter space under controlled conditions. Notably, see \citet{Barnes_2018_article} (hereafter B18), \citet{Ntampaka_2019} and \citet{Su_2020} (hereafter S20). B18 draws a general portait of the galaxy cluster population within the IllustrisTNG simulations. They examine several classification criteria for CC clusters and look at correlations between the different properties as well as their evolution with redshift. They find that the different classification criteria do not agree over the entire cluster population, and that the correlations between different criteria are dependent on cluster mass. They also note that at low redshift (z$<$0.25), the CC fraction in TNG is lower than in observations for 4 of their 6 criteria.\\
\\
Given the need for precise analysis to decode the diverse properties of the ICM in observations and simulations, traditional analytical methods can be augmented by contemporary approaches. In this context, machine learning tools have emerged as a powerful ally. Over the last decade, their application has grown in popularity, transforming how we analyze and interpret astronomical data. Machine learning algorithms have been successfully applied to a variety of tasks in galaxy cluster analysis, for example to estimate galaxy cluster masses (\citealt{Ntampaka_2019}, \citealt{Andres_est_masses}, \citealt{Krippendorf}) and mass accretion rates (\citealt{soltis_2024}), to infer galaxy cluster mass profiles (\citealt{Ferragamo_mass_profiles}), to find galaxy cluster members (\citealt{Angora_cluster_members}), or to classify galaxy clusters into CC, WCC and NCC (\citealt{Su_2020}).\\
\\
Notably, \citet{Ntampaka_2019} use a convolutional neural network (CNN) to estimate galaxy clusters masses from simulated \textit{Chandra} X-ray images obtained from IllustrisTNG clusters. Their results show that a low resolution image (128 $\times$ 128 pixels) can be successfully used to predict cluster mass with a reasonably low scatter (12\%). They also perform an interpretability analysis with Google DeepDream tools, allowing to see which parts of the cluster image are most significant for the prediction. \\
\\
On the other hand, S20 tackle a classification problem. They use a residual network (ResNet; \citealt{resnet_paper}) with convolutional layers to classify mock \textit{Chandra} images of IllustrisTNG galaxy clusters into CC, WCC and NCC clusters. They use the radiative cooling time to define their class labels, and obtain an average balanced accuracy (BAcc) of 0.85. Using class activation maps, they are able to locate which regions are discriminative for the network. Their results show that an X-ray image of a galaxy cluster contains enough information for a neural network to classify the cluster as CC, WCC or NCC when using the cooling time as the classification criteria.\\
\\
We expect to learn more about galaxy clusters in the following years from the X-ray telescope eROSITA (extended ROentgen Survey with an Imaging Telescope Array). Built by the Max Planck Institute for Extraterrestrial Physics (MPE), it is the primary instrument on the Russian-German `Spectrum-Roentgen-Gamma' (SRG) mission. Launched in 2019, it is currently performing an all-sky survey within an energy range of 0.2 - 10 keV and it is expected to detect at least 100 000 massive galaxy clusters (\citealt{Merloni_2012_erosita}, \citealt{Pillepich_2012_erosita}). With this large amount of data to analyze comes the need for an efficient, fast, and automated way to characterize the ICM in galaxy clusters is even more important. \\

Therefore, in this work, we continue the analysis by exploring how other galaxy cluster classification metrics behave under machine learning analysis. We present a deep learning approach for galaxy cluster characterization. We analyze five key properties of the hot gas in galaxy clusters, often used as classification metrics for their cooling properties: the central cooling time, the central electron density, the central entropy excess, the concentration parameter, and the cuspiness parameter. We select these properties as they are the most widely used in the literature for galaxy cluster classification (e.g. \citealt{cavagnolo_2009_article}, \citealt{santos_2010}, \citealt{Santos_2008}, \citealt{Hudson_2010_article}, \citealt{Andrade_Santos_2017}, \citealt{Barnes_2018_article}, \citealt{mcdonald_2019}). However, as hinted in \citet{Barnes_2018_article}, the classification results from the five metrics do not always agree, and the choice of the best metric to use is subject to debate. \\
\\
Our work therefore aims to compare the values of the five metrics, analyzing how informative they are and how well they lend themselves to machine learning analysis. We construct a neural network that simultaneously predicts the values of all five cluster properties from mock X-ray images; we then analyze the predictions using simulation-based inference. The paper is structured as follows. In Section \ref{sec:simulated_data}, we describe the IllustrisTNG simulations, the mock \textit{Chandra} images as well as the input preprocessing steps. We present the five chosen classification metrics in Section \ref{sec:metrics}.  In Section \ref{sec:clustering}, we describe our approach for the clustering algorithm and present corresponding results. In Section \ref{sec:regression}, we present our network architecture for the regression task as well as results. In Section \ref{sec:discussion}, we present our simulation-based inference (SBI) approach to obtain posterior probability distributions for our network predictions and we discuss our results. We conclude in Section \ref{sec:conclu}.\\
\section{Simulated Data and Metrics} \label{sec:methods}
It is known that training a supervised machine learning algorithm requires large amounts of labeled data. Since observational datasets of galaxy clusters for which the five classification metrics have been calculated are very limited, we turn towards simulations to provide a larger dataset. We select the IllustrisTNG simulations as the large volumes (300 comoving Mpc) provided by the simulations yield a large number of massive galaxy clusters to analyze.\\
\subsection{Simulated Data} \label{sec:simulated_data}

\subsubsection{IllustrisTNG Clusters}

We use galaxy cluster data from the cosmological gravo-magnetohydrodynamical simulations IllustrisTNG (\citealt{illustris_release_paper}, \citealt{illustris_1_pillepich}, \citealt{illustris_2_springel}, \citealt{illustris_3_nelson}, \citealt{illustris_4_naiman}, \citealt{illustris_5_marinacci}), run with the \textsc{arepo} (\citealt{Springel_2010_article}) dynamical mesh code. The simulations use a $\Lambda$CDM model for the background cosmology, assuming cosmological constants from the Planck collaboration results: $H_0=68$ km s$^{-1}$ Mpc$^{-1}$, $\Omega_M=0.31$, $\Omega_{\Lambda}=0.69$ (\citealt{planck_2016}). The simulations are made up of three volumes: TNG50, TNG100 and TNG300; the number refers to the size of the simulation box in units of comoving Mpc. The large volumes as well as high resolution ($\lesssim$ 1 kpc scales; \citealt{illustris_release_paper}) provided by these simulations make them suitable for studying the gas in galaxy clusters. Each simulation includes dark matter, gas, stars and supermassive black holes, solving for the coupled evolution of all components through time, with a redshift range of $z=127$ to $z=0$.\\
\\
We select clusters with a mass  $M_{500c}> 10^{13.57} M_{\odot}$ from the $z=0$ snapshot of the TNG300 simulation, similar to the work in \citet{Ntampaka_2019}. We restrict our training data to a single redshift to evaluate the model's performance without the confounding factor of cluster evolution through time.  We select the  z=0 snapshot because it has the largest number of massive clusters.
Haloes are identified using the Friends-of-Friends (FoF) algorithm, run on the dark matter particles. This results in 606 clusters, with an $R_{500c}$ range of 517 kpc to 1624 kpc. The value for $R_{500c}$ is taken directly from the IllustrisTNG FoF catalog, using the variable \texttt{Group\_R\_Crit500} and $h=0.68$.\\
\\
\subsubsection{Center of the cluster}
\label{sec:find_center}

\begin{figure}
\includegraphics[width=0.42\textwidth]{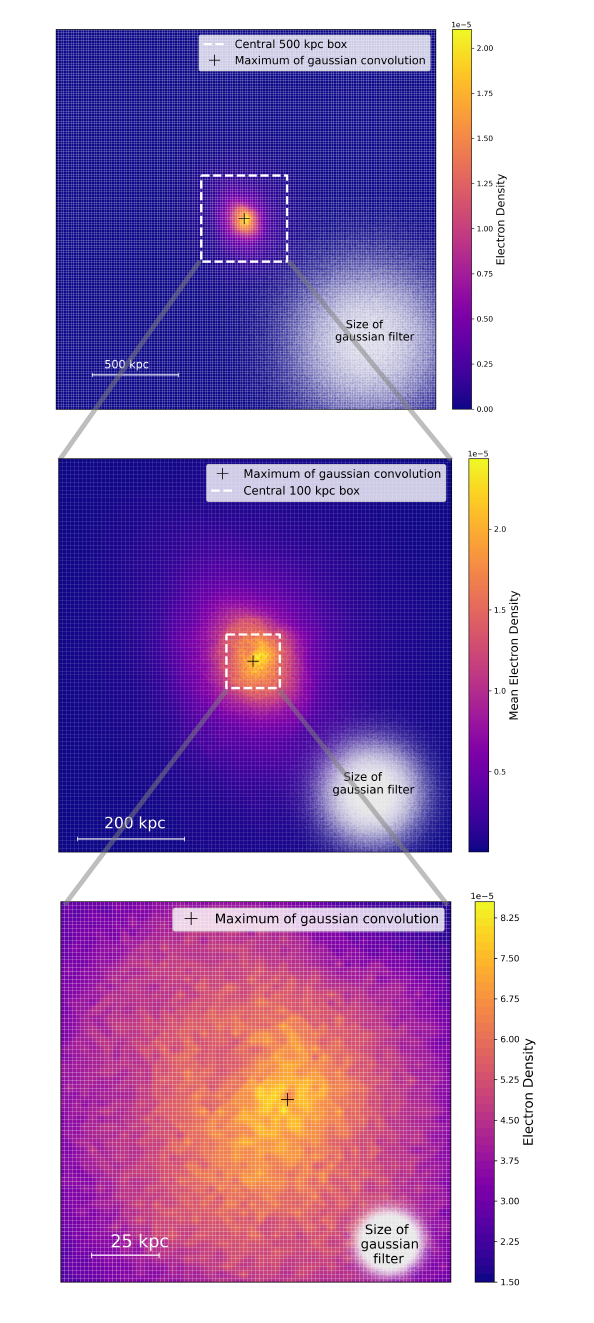}
  \caption{Images corresponding to the three steps of the iterative process to define the cluster center described in Section \ref{sec:find_center}. For each image, we first plot a two-dimensional histogram of the electron density. We then overlay a white grid representing the binning of the data as well as a white cloud (at the bottom right), representing the size of the gaussian filter used in the convlution during that step. The black cross corresponds to the maximum point of the gaussian convolution. The white square represents the central region which served as the starting point for the next step, as indicated by the grey lines. This iterative process allows us to examine both large-scale and small-scale aspects of the electron density distribution, ultimately providing a robust estimate of the X-ray peak.}
  \label{fig:iter_pdf}
\end{figure}

Galaxy cluster centers are needed to calculate gas properties and generate mock X-ray emission, but locating them is a complex task. \\
\\
There are numerous ways of defining the center of a galaxy cluster, such as the location of the BCG, the minimum of the gravitational potential, the X-ray/optical centroid or the X-ray/optical peak. These definitions result in centers that are offset from one another, in observational data (e.g. \citealt{McDonald_2014}), as well as in simulated data (\citealt{Cui_2016}). \\
\\
Since we are working with X-ray data and are interested in the morphology of the cluster, we experiment with the X-ray large-scale centroid as well as the X-ray peak. The large-scale centroid is an appropriate choice for NCC clusters, as the X-ray emission is more or less uniform. However, for CC clusters, the core is often offset from the large-scale centroid, as the TNG clusters are not perfectly spherical. This is also the case for real galaxy clusters (e.g. \citealt{Lazzati_1998}). Since we are interested in locating the core for these clusters, and therefore  define the center of the cluster as the X-ray peak. We found that the calculated metrics (Section \ref{sec:metrics}) for NCC clusters were not affected by this choice since the gas distribution is more uniform.\\
\\
To find the X-ray peak, we look at the smoothed 3D distribution of the gas density. We only consider the gas cells for which all of the emission lies in the X-ray range; we therefore select gas cells with a temperature above $10^7$ K.  We wish to see where the data most resembles a 3D gaussian peak, and select that point to represent the central peak of the cluster, with a precision of 1 kpc. To do this, we apply a gaussian filter to the data with $\sigma=$5 kpc, where $\sigma$ is the standard deviation of the gaussian kernel. Since convolving a 3D kernel with a precision of 1 kpc over the entire cluster would be too time-consuming, we use an iterative approach, starting with a coarse grid and gradually transitioning to a finer grid. We first bin the data into 10 kpc cubes, and convolve the binned data with a gaussian filter that has $\sigma$=250 kpc. The 500 kpc box around the maximum point of this convolution corresponds to a rough estimate of the central region. We then select the cells of this region and repeat the process: we bin the data into 5 kpc cubes and convolve it with a $\sigma$=50 kpc gaussian filter. Finally, we select the cells within a 100 kpc box around the new maximum and repeat the process a third time with 1 kpc bins and $\sigma$=5 kpc. We define the maximum point of the last convolution as the location of the peak of the gas density, corresponding to the 3D cluster center. This process is illustrated in figure \ref{fig:iter_pdf}.\\

\subsubsection{Mock \textit{Chandra} Observations}
We then generate mock X-ray observations of the clusters with the \texttt{pyXSIM} (version 3.0.1) and \texttt{SOXS} (version 3.0.2) packages (\citealt{ZuHone_2014}). The \texttt{pyXSIM} package is an implementation of the \texttt{PHOX} algorithm (\citealt{Biffi_2012}, \citealt{Biffi_2013}), and it simulates X-ray emission from astrophysical sources. Using the gas properties extracted from the IllustrisTNG data for each cluster, such as density, temperature and metallicity, photons are generated in a 3D space with the \textsc{apec} emission model (\citealt{apec_model}). We specify the calculated 3D center (see Section \ref{sec:find_center}) as the center of the spherical region from which the photons are generated, and we assume a redshift of $z=0.05$, as done in \citet{Ntampaka_2019} and S20. The photons are then projected along the three axes of the simulation box to create three separate images. We chose to only take these 3 projections along the perpendicular axes, as a compromise between data augmentation and ensuring that the different samples remain as independent as possible. We then apply foreground galactic absorption with the \texttt{tbabs} model (\citealt{tbabs_model}), assuming a typical value of $N_H$ = 0.04 $ \times 10^{22}$ cm$^{-2}$ for the column density (as in e.g. \citealt{Ntampaka_2019}, S20).\\
\\
The simulated photons landing on the detector are then convolved with an instrumental response matrix using the \texttt{SOXS} package. This package creates simulated X-ray observations of X-ray sources, and has numerous simulated instruments available including realistic instrumental noise. We select \textit{Chandra}'s ACIS-I detector, as its large field of view is well-suited for observing low-z clusters. We select the configuration corresponding to the Cycle 0. We integrate for a typical exposure time of 100.0 ks, which ensures that we detect enough photons. The final mock images display the soft-band X-ray emission (between 0.5 and 2.0 keV) typical of the ICM in galaxy clusters.

\subsubsection{Image preprocessing}
\label{sec:preprocessing}

As a preprocessing step, we use a min-max normalization ($\frac{\text{value} - \text{min}}{\text{max} - \text{min}}$) for every image, constraining the value of the pixels between 0 and 1 so that all the inputs have the same scale. However, the maximum point of the mock X-ray image is sometimes a point source in the background, which is not informative for the cluster morphology. Therefore, we first remove point sources from our mock \textit{Chandra} images, using the CIAO 4.15 software package (\citealt{ciao_paper}). We locate the point source regions with \texttt{wavdetect}, and then use \texttt{dmfilth} to replace the source pixels with values interpolated from background regions. Once the point sources have been removed, we verify manually that the brightest pixel in each image actually belongs to the galaxy cluster in the X-ray image and not a background point source.\\
\\
In order to reduce the dimension of the input, we downsample the images to a size of 256 $\times$ 256 pixels by applying 8 $\times$ 8 binning. We then apply a min-max normalization to each image. Finally, the dataset is augmented with flips along the vertical and horizontal axes as well as with 90$^{\circ}$ rotations, which results in 8 different views of the same image. Considering the 2D projections along the three axes, we end up with 24 images per cluster and a total of 14 544 images for our dataset.

\subsection{Galaxy cluster classification metrics}
\label{sec:metrics}
In order to characterize the galaxy clusters,  we have chosen five classification metrics: the central cooling time, the central electron density, the entropy excess, the cuspiness parameter and the concentration parameter. The concentration parameter depends only on information contained within the mock X-ray image. The four other metrics require computing density and temperature profiles. In our case, we use the information directly available within the IllustrisTNG gas cells to compute these profiles and calculate the metrics. The detailed process is described in the following paragraphs.\\
\\

When observing galaxy clusters in the X-ray range, the pattern on the detector is a projection of all the photons along the line of sight, since the gas is optically thin. In an effort to mimic observations, we use projected 2D values to calculate the metrics of interest. We project our data along the three axes of the simulation box, and compute 2D centers for the three projections of each cluster.
We decided to recalculate a 2D center for each projection of the same cluster, rather than using the same 3D center for all three projections. This was done in an effort to make the projections of a same cluster more independent from each other, in addition to mimicing observations. We use the same iterative method described in Section \ref{sec:find_center} to compute the center, but now convolve a 2D gaussian filter over the projected 2D distribution of the gas density. The values for bin width and for $\sigma$ stay the same as above. We then construct radial projected profiles for the electron density and the temperature of the gas, using 50 radial logarithmic bins in the range $10^{-3}$ - 1.5 $R_{500}$, as done in B18. For each bin, we take a weighted mean of the observed value (the electron density or the temperature) for the gas cells within that annulus. The value is weighted by the emission measure, which is defined as the square of the electron density. We use the weighted standard deviation as the uncertainty for each bin.\\
\\
Since we are calculating projected 2D radial profiles, we end up with different observed values for the three different projections of the same cluster, and therefore different values for the metrics. Since we have 606 initial clusters, accounting for these three different projections of each cluster, we obtain 1818 ($3\times606$) observations when talking about the distribution of the metrics, which we will discuss in more detail in the following sections.\\

\begin{figure}
  \centering
  {\includegraphics[width=0.4\textwidth]{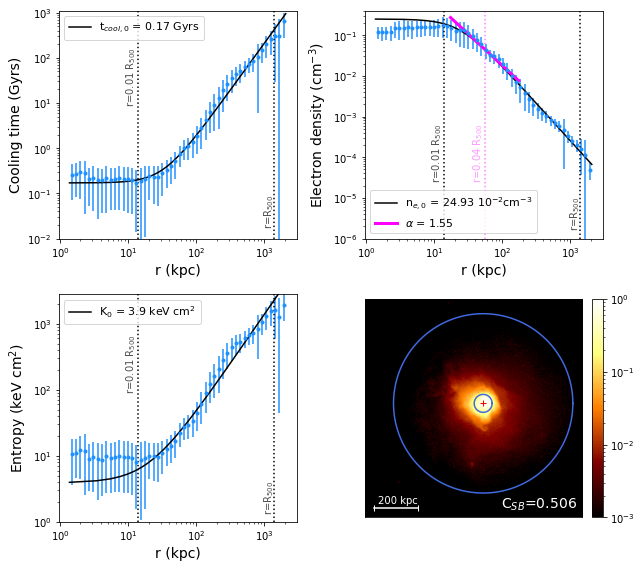}}\\
  \caption{Example of cluster properties for a cool core cluster. In the top left, is shown the 2D cooling time profile. The data points are in blue, the best-fit model is in black and the fitted values are shown in the legend. The black dotted vertical lines at 0.01 $R_{500}$ and $R_{500}$ represent the boundaries for the fit, as dicussed in Sections \ref{sec:cooling_time} through \ref{sec:central_K0}. In the top right, is shown the 2D electron density profile, as well as the tangent line in pink that is used to calculate the cuspiness at $r=0.04 R_{500}$ (see \ref{sec:cuspiness}). The fit value for $n_{e,0}$ and the cuspiness $\alpha$ are shown in the legend as well. In the bottom left, is shown the 2D entropy profile, with the fit value for $K_0$ shown in the legend. In the bottom right, is shown the preprocessed (binned and normalized) 2D X-ray image  that is used as input for the neural network. The red cross represents the cluster center, identified using the method in Section \ref{sec:find_center}, and the blue circles at $r=40$ kpc and $r=400$ kpc represent the regions used in the calculation of the concentration parameter $C_{SB}$ (see \ref{sec:conc}). The value for $C_{SB}$ is shown in the bottom right. The colorbar for normalized the X-ray image is shown in log scale. The profiles shown in these figures are representative of the cool core clusters found in our dataset.}
  \label{fig:CC_properties_img}
\end{figure}

\begin{figure}
  \centering
\includegraphics[width=0.4\textwidth]{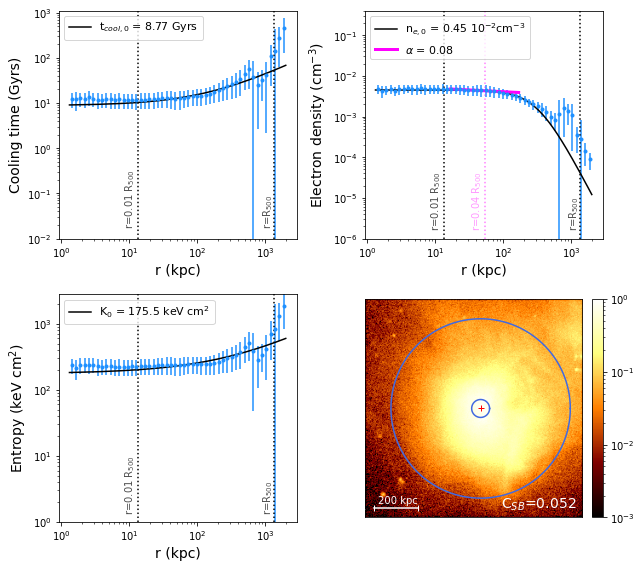}
  \caption{Same as figure \ref{fig:CC_properties_img}, but for a non-cool core cluster. For each property, we use the same y axes as in figure \ref{fig:CC_properties_img}. This allows us to clearly see the differences in the profiles between a cool-core cluster and a non-cool core cluster. We notice that for a non-cool core cluster, the profiles (for all shown physical properties) are more uniform than for a cool core cluster; the difference is much less pronounced between the core of the cluster and the outskirts. This can be explained by the uniform distribution of the ICM in a non-cool core cluster.}
  \label{fig:NCC_properties}
\end{figure}

\subsubsection{Central cooling time}
\label{sec:cooling_time}
The central cooling time is a widely used metric (\citealt{Hudson_2010_article}, \citealt{McDonald_2013_article}) to classify galaxy clusters into CC, WCC and NCC. It is defined as

\begin{equation}
\label{eqn:cooling_time}
    t_{cool} = \frac{3 (n_e + n_p)kT}{2 n_en_H \Lambda(T,Z)}
\vspace{5mm}
\end{equation}
where $\Lambda$ is the cooling function for an optically thin plasma (\citealt{Sutherland_Dopita_1993}). We assume $n_p = 0.92 n_e$ and $n_H = 0.83 n_e$, as done in \citet{McDonald_2018_np_nh}. We  use the cooling function table for a metallicity of [Fe/H]=-0.5 from \citet{Sutherland_Dopita_1993}, interpolating $\text{log}(T)$ to obtain $\text{log}(\Lambda)$. \\
\\
Rather than only considering the gas cells at the center of the cluster to compute the central cooling time, we opt to use the entire cooling time profile and fit for the central value, especially since the TNG data tends to be noisy and often have multiple small overdensities near the center that could bias our estimate. We use our 50-bin density and temperature radial profiles to obtain a cooling time profile within a range of $10^{-3}$ to 1.5 $R_{500}$. We also propagate uncertainties through equation \ref{eqn:cooling_time}. We follow the method used in B18 and \citet{cavagnolo_2009_article} to calculate the central entropy excess, and expand the approach to the cooling time. More precisely, we fit the cooling time in the range $0.01 - 1.0 R_{500}$ with a power-law of the form:

\begin{equation}
    t_{\text{cool}}(r) = t_{\text{cool},0} + t_{\text{cool}, 100}\left(\frac{r}{100 \text{kpc}}\right)^{\gamma},
\vspace{5mm}
\end{equation}
where $t_{\text{cool}, 0}$ is the central cooling time, $t_{\text{cool}, 100}$ is a normalization factor at 100 kpc, $r$ is the radial distance from the center and $\gamma$ is the power-law index (\citealt{cavagnolo_2009_article}). We fit this model to our data to get a value for the central cooling time $t_{\text{cool}, 0}$. We do this with a Markov-Chain Monte Carlo (MCMC) with uniform priors, using the \texttt{emcee} python package (\citealt{emcee_paper}). We use 100 walkers and run the MCMC for 10 000 steps. We also perform sigma clipping to remove outliers (e.g. small overdensities or numerical artefacts) outside of 3 $\sigma$ prior to fitting, where $\sigma$ is the standard deviation of the $t_{\text{cool}}$ values. We use the $t_{\text{cool}, 0}$ parameter as the classification metric, where clusters with $t_{\text{cool}, 0}$ $<$ 1 Gyr are defined as CCs, those with 1 Gyr $<$ $t_{\text{cool}, 0}$ $<$ 7.7 Gyr are WCCs, and those with $t_{\text{cool}, 0}$ $>$ 7.7 Gyr are NCCs (\citealt{Hudson_2010_article}, \citealt{McDonald_2013_article}, B18). Out of our 1818 observations, we have 50 CCs, 908 WCCs and 860 NCCs with these thresholds. The distribution of central cooling time values is shown in the first panel of figure \ref{fig:metrics_distribution}.\\
\\
\subsubsection{Central electron density}
\label{sec:central_ne}
\begin{figure}

\includegraphics[width=0.4\textwidth]{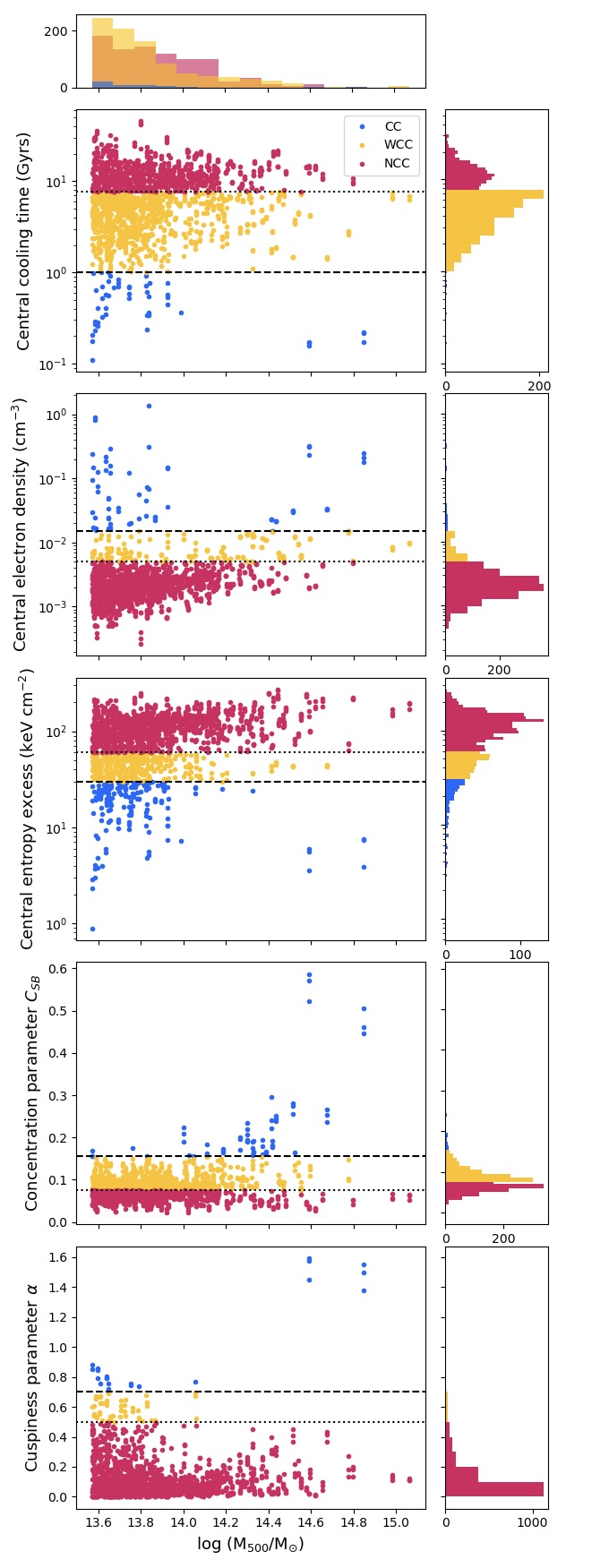}
\caption{ Distribution of cluster properties (central cooling time, central electron density, central entropy excess, concentration and cuspiness) as a function of the cluster mass for the 1818 observations in our sample. The CCs for each metric are plotted in blue, the WCCs are in yellow and the NCCs are in red. The colors in the mass histogram at the top are based on the central cooling time, as it is the most commonly used metric. We notice that the different classification metrics don't always agree, as the propotions of CCs, WCCs and NCCs are not the same for all metrics. We also notice that NCCs are over-represented in our sample.} \label{fig:metrics_distribution}
\end{figure}

For the central electron density, we use the same method as for the central cooling time (described in \ref{sec:cooling_time}) to construct the radial profile that we fit with a MCMC. We use the 50-bin density radial profile with a range of $10^{-3}$ - 1.5 $R_{500}$, and fit it in the range 0.01 - 1.0 $R_{500}$ using the following equation:

\begin{equation}
    n_e(r) = n_{e,0} \left(1+\frac{r^2}{r_c^2}\right)^{-3\beta/2},
    \label{eqn:density}
\vspace{5mm}
\end{equation}
where $n_e(r)$ is the electron density, $r$ is the radial distance from the center, $r_c$ is the core radius, $n_{e,0}$ is the central value of the electron density, and $\beta$ is a power-law index (\citealt{Cavaliere_1976}, \citealt{cavaliere_1978}, \citealt{Grego_2001}, \citealt{Reese_2002}). We chose this model because it is convenient to work with, and after examination it seemed to correspond well to our data. Alternative choices for the electron density model are explxored in  the discussion in Section \ref{sec:metric_comp}. We use a MCMC to fit the model to our data, and use the $n_{e,0}$ parameter as a the metric, where clusters with $n_{e,0} >$  1.5 $\times$ $10^{-2} \text{cm}^{-3}$ are defined as CCs, those with 0.5 $\times$ $10^{-2} \text{cm}^{-3}$  $< n_{e,0} <$  1.5 $\times$ $10^{-2} \text{cm}^{-3}$ are WCCs, and those with $n_{e,0} <$ 0.5 $\times$ $10^{-2} \text{cm}^{-3}$ are NCCs (\citealt{Hudson_2010_article}, B18). Out of our 1818 observations, we have 68 CCs, 175 WCCs and 1575 NCCs with these thresholds. The distribution of central electron density values is shown in the second panel of figure \ref{fig:metrics_distribution}.\\

\subsubsection{Central entropy excess}
\label{sec:central_K0}
The entropy $K$ is defined as 
\begin{equation}
    K = \frac{k_BT}{n_e^{2/3}},
    \label{eqn:entropy}
\vspace{5mm}
\end{equation}
where $k_B$ is the Boltzmann constant, $T$ is the temperature and $n_e$ is the electron density. Once again, following the method described in \ref{sec:cooling_time}, we use the 50-bin density and temperature radial profiles to obtain an entropy radial profile with a range of $10^{-3}$ - 1.5 $R_{500}$, and we propagate uncertainties through equation \ref{eqn:entropy}. We fit the entropy $K$ in the range 0.01 - 1.0  $R_{500}$ with a power-law of the form:\\

\begin{equation}
    K (r) = K_0 + K_{100} \left(\frac{r}{100 \text{kpc}}\right)^{\gamma},
\vspace{5mm}
\end{equation}

where $r$ is the radial distance from the center, $K_{100}$ is a normalizatoin factor at 100 kpc, $\gamma$ is a power-law index, $K_0$ is the central entropy excess, defined as the excess entropy above the best-fitting power-law at larger radius (\citealt{cavagnolo_2009_article}). We use a MCMC to fit the model to our data, and use the $K_0$ parameter as the metric. Clusters with $K_0 <$ 30 keV cm$^2$ are defined as CCs, those with 30 keV cm$^2$ $ < K_0 < $ 60 keV cm$^2$ are WCCs, and those with $K_0 > $  60 keV cm$^2$ are NCCs (B18). Out of our 1818 observations, we have 185 CCs, 390 WCCs and 1243 NCCs with these thresholds. The distribution of central entropy excess values is shown in the third panel (from the top) of figure \ref{fig:metrics_distribution}.\\
\\
\subsubsection{Concentration parameter}
\label{sec:conc}
Introduced by \citet{Santos_2008}, the X-ray concentration parameter is the ratio of the X-ray luminosity within the core compared to the luminosity within a larger radius:

\begin{equation}
    C_{SB} = \frac{F (r<40\text{kpc})}{F (r<400 \text{kpc})},
\vspace{5mm}
\end{equation}
where $C_{SB}$ is the concentration parameter (the subscript SB refers to the surface brightness), $F$ is the X-ray flux, and $r$ is the projected radial distance from the center. We compute $C_{SB}$ by taking the ratio of photons within 40 kpc and within 400 kpc in the mock X-ray images, as show in the bottom right panel of figures \ref{fig:CC_properties_img} and \ref{fig:NCC_properties}. $C_{SB}$ is a dimensionless number between 0 and 1, and clusters with $C_{SB}$ $>$ 0.155 are defined as CCs, those with 0.075 $< C_{SB}< $ 0.155 are WCCs, and those with $C_{SB} < $ 0.075 are NCCs (\citealt{Santos_2008}). Out of our 1818 observations, we have 57 CCs, 858 WCCs and 903 NCCs with these thresholds. The distribution of concentration values is shown in the fourth panel of figure \ref{fig:metrics_distribution}.\\
\\
\subsubsection{Cuspiness}
\label{sec:cuspiness}
Introduced by \citet{Vikhlinin_2007}, the cuspiness parameter $\alpha$ aims to quantify the importance of the central cusp in the X-ray brightness distribution, and is defined as:

\begin{equation}
    \alpha = - \left. \frac{\text{d log }n_e(r)}{\text{d log } r}\right|_{r = 0.04 R_{500}},
\vspace{5mm}
\end{equation}
where $r$ is the radial distance from the center and $n_e$ is the electron density. Since we have a theoretical profile for the density (equation \ref{eqn:density} with parameters $n_{e,0}$, $\beta$ and $r_c$ estimated by the MCMC), we can compute the derivative $\frac{\text{d log }n_e(r)}{\text{d log } r}$, evaluated at $r=0.04 R_{500}$. We then use $\alpha$ as the metric, where clusters with $\alpha$ $>$ 0.7 are defined as CCs, those with 0.5 $<$ $\alpha$ $<$ 0.7 are WCCs, and those with $\alpha$ $<$ 0.5 are NCCs (\citealt{Vikhlinin_2007}) Out of our 1818 observations, we have 21 CCs, 52 WCCs and 1745 NCCs with these thresholds. The distribution of cuspiness values is shown in the fifth panel of figure \ref{fig:metrics_distribution}.\\

\subsubsection{Comparison of the metrics}
\label{sec:methods_comparison}

An example of all five cluster properties is shown in figure \ref{fig:CC_properties_img} for a CC cluster and figure \ref{fig:NCC_properties} for a NCC cluster. These specific clusters were chosen as all five metrics are in agreement for their classification. Since the y axis is the same for the plots on the top (CC) and the bottom (NCC), we can clearly see the difference in the profiles. The value of the cooling time is similar at large radius ($\approx$ $R_{500}$) for both clusters. For the CC cluster, the cooling time drops drastically towards the center of the cluster, then becomes flat at the core. The same can be observed for the entropy profile. For the density, the profile is again similar at large radii. The CC density profile then keeps increasing towards the center while the NCC profile remains flat. This is also seen through the value of the cuspiness $\alpha$, representing the slope of the density profile. Finally, the X-ray image shows the difference in the distribution of the X-ray brightness: for the CC the emission is concentrated at the center, while for the NCC the emission is more uniform. It is important to note that the X-ray images shown have been normalized (the maximum value is 1 for both), so the overall perceived brightness doesn't reflect the total emission of the cluster. \\

\begin{figure*}
    \centering
    \includegraphics[width=\textwidth]{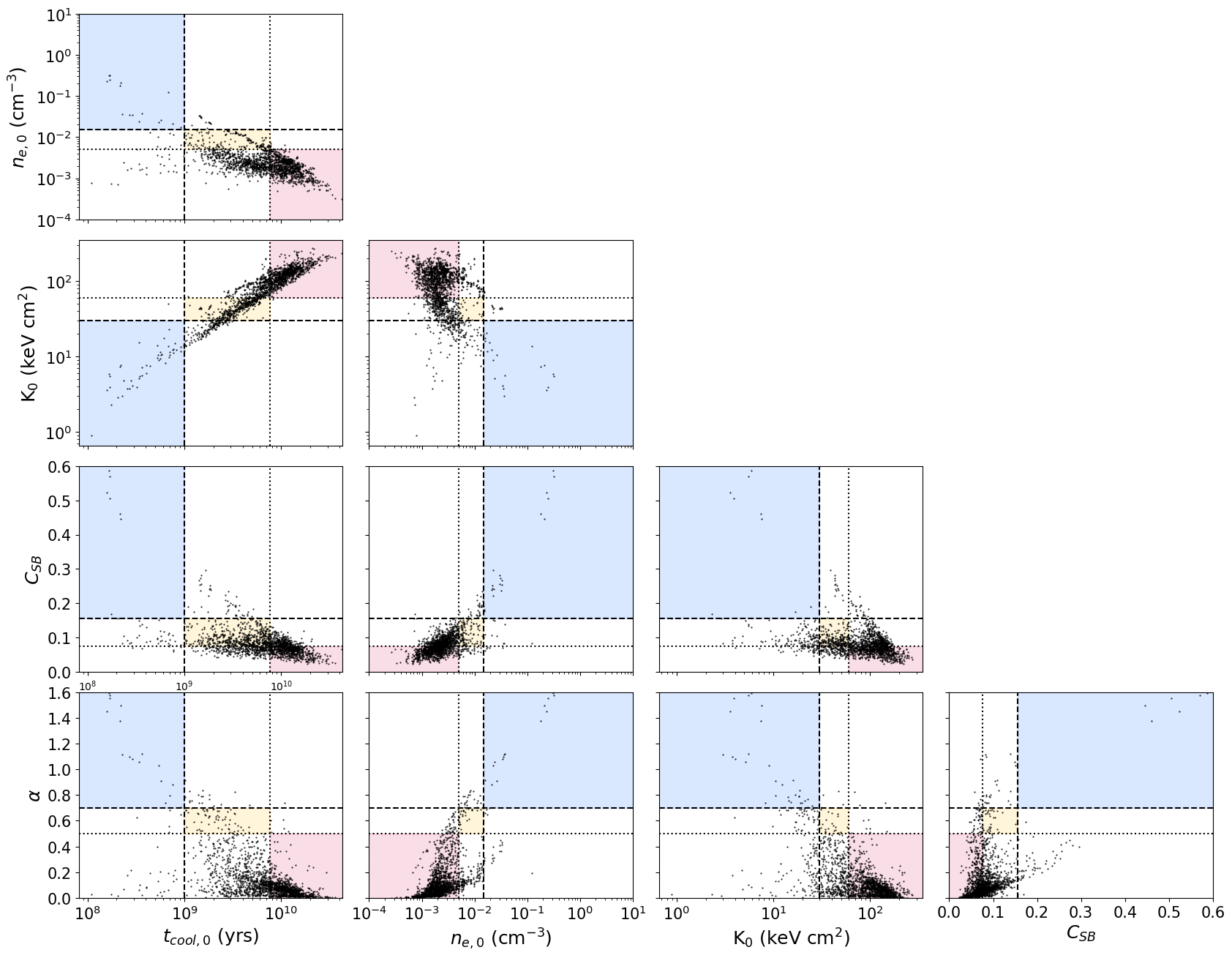}
    \caption{Comparison of the cluster properties for the 1818 observations in our sample. The shaded regions in each plot represent the zones where the clusters would belong to the same class according to both criteria (blue region for CCs, yellow for WCCs, red for NCCs). The dashed line represents the cutoff between CCs and WCCs, whereas the dotted line is the cutoff between WCCs and NCCs.  Similar to figure \ref{fig:metrics_distribution}, this figure highlights the disagreement between classification metrics, as well as the relative importance of clusters of each category (CC/WCC/NCC) in the dataset.}
    \label{fig:corner_plot}
\end{figure*}

The distribution of the five cluster properties as a function of mass is shown in figure \ref{fig:metrics_distribution}. The imbalance between the three classes can be clearly seen; while the exact proportion varies with the properties, there are significantly more NCCs than WCCs and CCs. We present a comparison of the five cluster properties in figure \ref{fig:corner_plot}. We observe a strong correlation  between the central cooling time and the central electron density (top left panel), as well as between the central cooling time and the central entropy excess. The correlations between the other metrics are not as significant, with a large proportion of data points not falling into the zones of agreement (shaded regions). As the NCCs are more numerous, we have more data points in the red shaded regions.\\

\section{Clustering}
\label{sec:clustering}
\subsection{Methods}
We first explore our data with an unsupervised k-means clustering algorithm. We wish to gather the data into 3 groups, and then see if these 3 groups ressemble the CC/WCC/NCC classification for any of the five metrics. To do this, we use the non-augmented version of the dataset, consisting of 1818 images. We start with the normalized images, complete with all the preprocessing steps described in Section \ref{sec:preprocessing}. However, we want the algorithm to be able to consider the morphology of the cluster as the key factor, rather than just the total amount of light in the image (i.e. the sum of all the pixel values). Therefore, we divide each normalized image by the sum of all pixels in that image.  This creates the 256 $\times$ 256 images used as input for the next step. We then perform principal component analysis (PCA) as dimensionalty reduction to make the data easier to process for the clustering algorithm, as it is known that k-means works better on lower dimension data (e.g. \citealt{clustering_high_dim}). The 256 $\times$ 256 images (65 536 features) serve as input, and are reduced to 1200 components per image. The number of components is chosen to explain 95\% of the variance. We then use k-means clustering to gather the data into 3 groups.\\

\subsection{Results}
\label{subsec:clustering_results}
\begin{figure*}
    \centering
    \includegraphics[width=\textwidth]{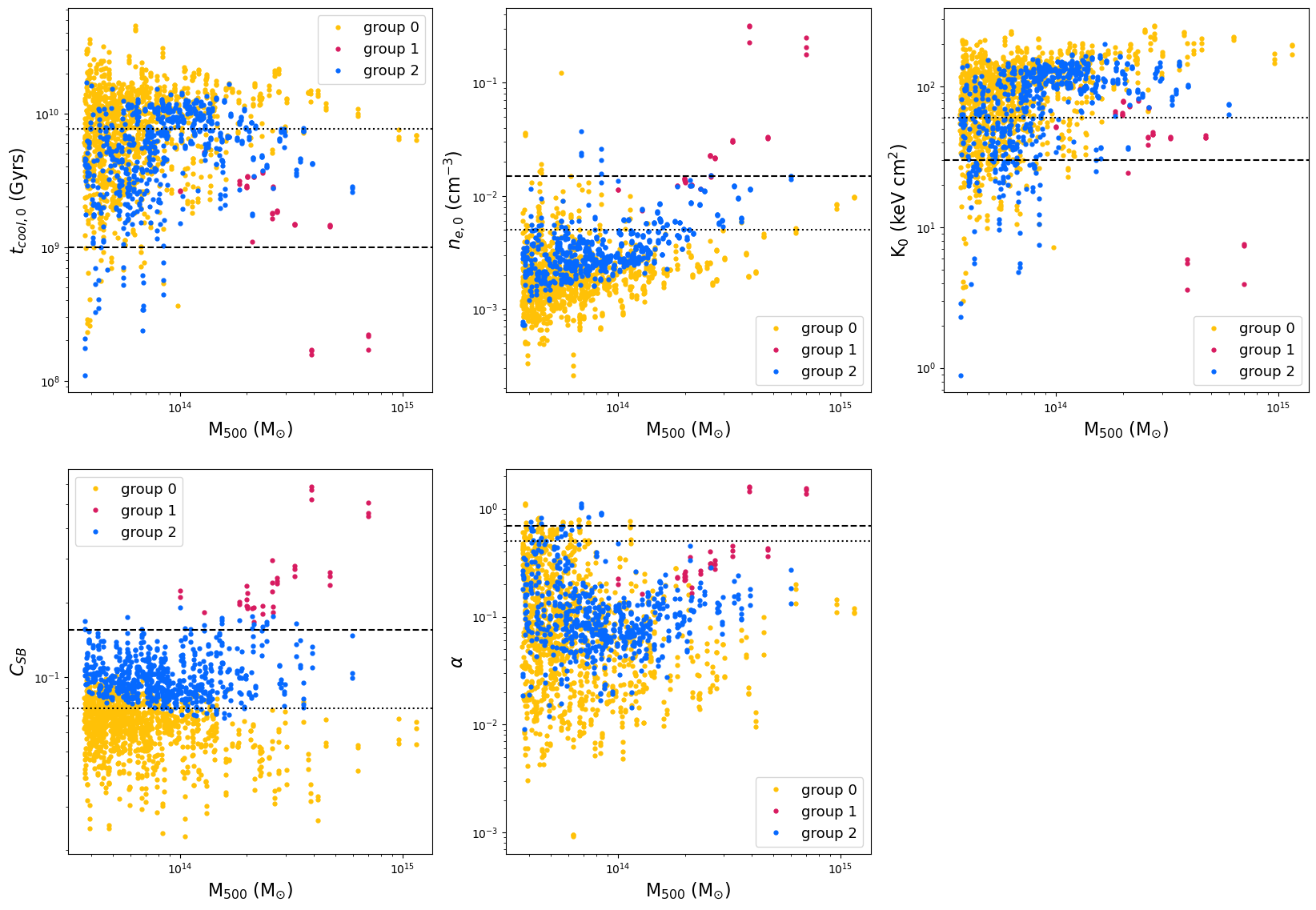}
    \caption{Results of the clustering algorithm. Each galaxy cluster property is plotted as a function of the cluster mass, with clusters belonging to group 0 in yellow, group 1 in red and group 2 in blue. The dashed line represents the cutoff between CCs and WCCs for each each metric; the dotted line represents the cutoff between WCCs and NCCs for each metric. As described in section \ref{subsec:clustering_results}, we notice a very strong correlation between the value of the concentration parameter $C_{SB}$ and the group index, while no such correlation is observed for the other four metrics.}
    \label{fig:clustering}
\end{figure*}

We present the results of the clustering algorithm in figure \ref{fig:clustering}. The galaxy clusters belonging to each of the three groups are shown in different colors. We can very clearly see that there is a strong correlation between the value of the concentration parameter $C_{SB}$ and the group index, with the point of transition between groups corresponding almost exactly to the CC and WCC cutoffs. Since the concentration is the only metric calculated directly from the X-ray image, it is expected to have the strongest correlation out of all the metrics. This result suggests that the PCA and the k-means clustering are able to isolate the shape of the X-ray brithgness profile (and therefore the morphology of the cluster and cool coreness) as a feature. The four other metrics do not exhibit such a correlation with the group index.\\

\section{Regression}
\label{sec:regression}

\subsection{Methods}

For the regression task, we have selected a ResNet model as our network architecture. This model utilizes convolutional layers, which are well suited for image analysis as they can extract meaningful spatial features and patterns in a translationally invariant way. ResNet models are made up of residual blocks, which allow the data to pass directly to a superior layer through skip connections. These models have been shown to achieve better results on various computer vision tasks with fewer parameters (\citealt{resnet_paper}), making them very efficient.\\

\begin{figure*}
    \centering
    \includegraphics[width=\textwidth]{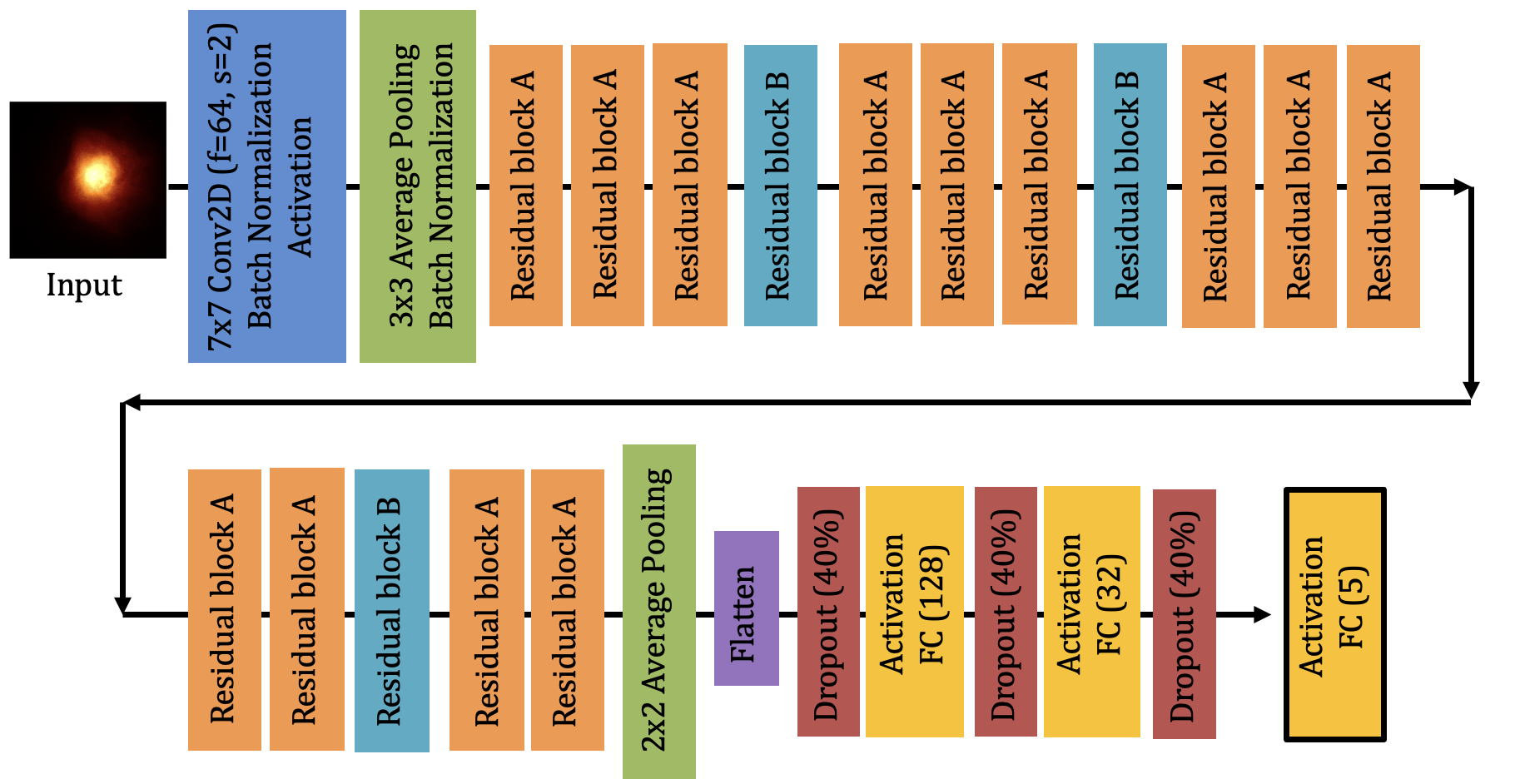}
    \caption{Architecture of our ResNet model used for the regression task. The input is a 256$\times$ 256 image. The residual blocks A and B are shown in figures \ref{fig:res_block_a} and \ref{fig:res_block_b} (in the Appendix). Conv2D is a 2D convolution, where $f$ and $s$ are the number of filters and the strides. FC is  a fully connected layer and the number in brackets corresponds to the number of neurons. The last layer [FC (5)] is the output layer, with the 5 neurons correponding to the 5 predictions.}
    \label{fig:my_resnet_model}
\end{figure*}

We use two types of residual blocks, as shown in the appendix, in figures \ref{fig:res_block_a} and \ref{fig:res_block_b}. In one case (block A), the input of the block is simply added two layers later through a skip connection. In the second case (block B), the skip connection involves a 1$\times$1 2D convolution layer. The full architecture is shown in figure \ref{fig:my_resnet_model}. The input (a 256 $\times$ 256 image) gets passed to a convolution layer, then a pooling layer, then a series of residual blocks A and B. The data is then pooled and flattened before getting passed to a fully-connected network with layers of 128, 32 and 5 neurons each, with 40\% dropout between each layer. The final 5-neuron layer is the output of the network, representing the prediction for log $t_{cool, 0}$, log $n_{e,0}$, log $K_0$, log $C_{SB}$ and log $\alpha$ (here `log' refers to log$_{10}$). We chose to predict the log of the values rather than the values themselves because this transformation spreads out the data in a way that amplifies the key differences and results in a range of values of the order of $10^0$ or $10^1$, making it easier to work with.\\

The activation function is a LeakyReLU with a negative slope of $\alpha=0.1$. We use the \textsc{Adam} optimizer with an initial learning rate of $10^{-4}$, dropping to $10^{-5}$ at the 400th epoch, and to $10^{-6}$ at the 700th epoch. The model is trained for 900 epochs. The loss is scaled so that cool core clusters are given more importance, since they are underrepresented in the sample. For each output we calculate a separate mean squared error loss, scaled by a factor of 1 to 5 depending on how close to a cool core the cluster is. For example, the loss for the smallest log $t_{cool,0}$ is scaled by 5, and the loss for the largest $t_{cool,0}$ is scaled by 1. The same is done for the other metrics with a factor of 5 corresponding to either the largest or the smallest value (depending on which end represents the CCs). The total loss is then calculated as the mean of the five scaled losses.\\

Our model takes as input 256 $\times$ 256 images, and is trained to predict a vector (log $t_{cool, 0}$, log $n_{e,0}$, log $K_0$, log $C_{SB}$, log $\alpha$) from these images. To create training, validation and test sets, the data is divided into folds, making sure that all 24 images of the same cluster always end up in the same fold. We first take 10\% of the clusters (61 out of 606 clusters, corresponding to 183 out of 1808 projections) for the test set. To ensure that it is representative, we verify that the CC/WCC/NCC proportion of the entire dataset for every metric is reflected in the test set. We then divide the remaining 545 clusters into 5 folds of 109 clusters each, again making sure that the CC/WCC/NCC proportion for every metric in each fold is about the same as for the entire dataset. We perform a 5-fold cross-validation, with 72\% of the data used for training, 18\% for validation, and 10\% for testing. Each model is trained 5 times with a different fold serving as the validation dataset each time. The hyperparameters that lend the best performance on all five folds are selected and then used to re-train a model using all five folds (90\% of the data) as the training dataset. This model is then evaluated on the test set.\\

\subsection{Results}

\begin{figure*}
    \centering
    \includegraphics[width=\textwidth]{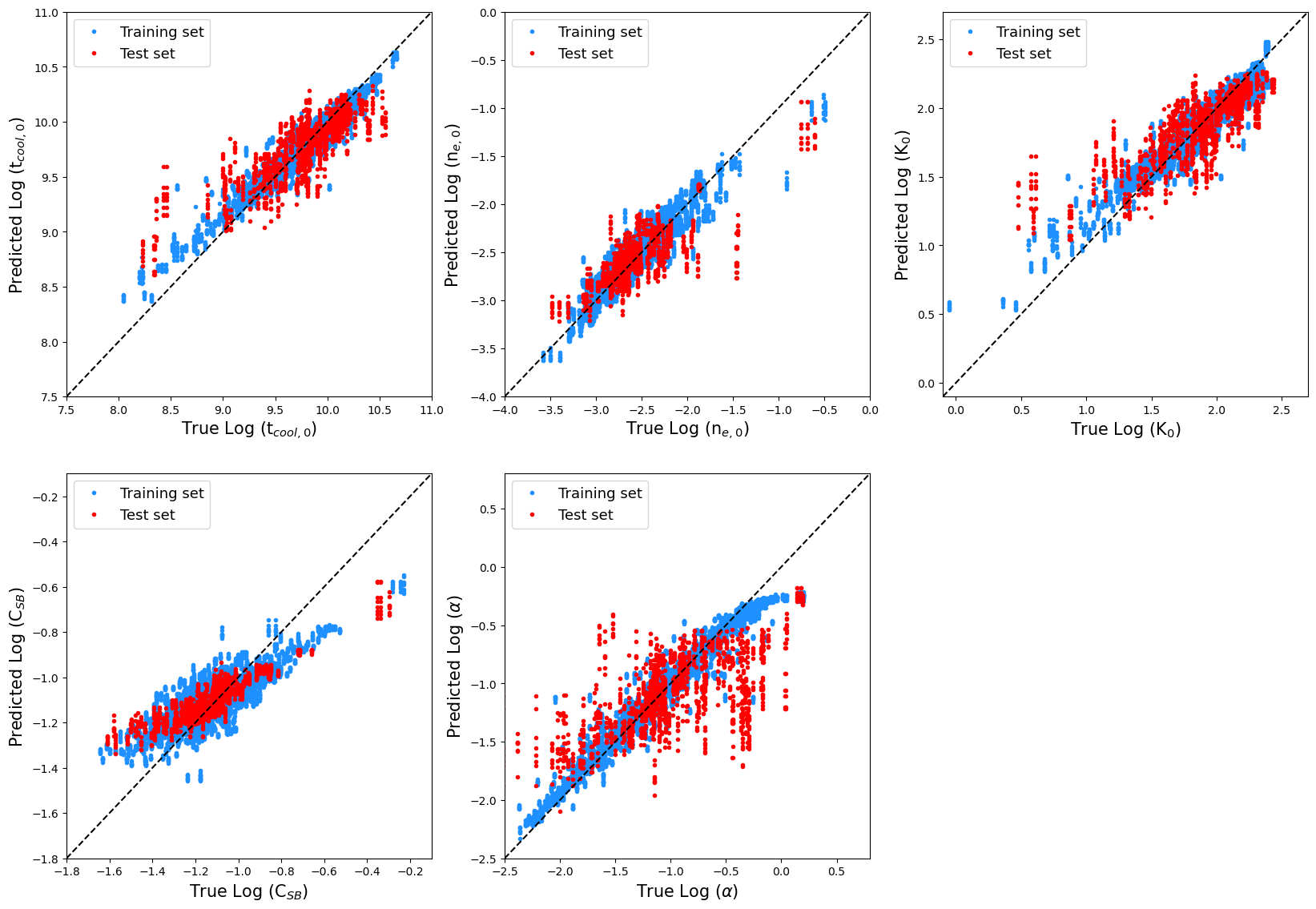}
    \caption{Predicted values as a function of true values for the central cooling time, the central electron density, the central entropy excess, the concentration and the cuspiness. The training data is shown in blue and the test data is shown in red. The vertical streams show the estimates for each of the 8 views of the same observation. We notice that for first four metrics, the predictions made on the test set are consistent with the ones made on the training set. For the cuspiness ($\alpha$), we notice that the network tends to overfit: the performance is good for the training set but unsatisfactory for the test set. This is analyzed further in the text. }
    \label{fig:preds_vs_true}
\end{figure*}

The predictions for all five cluster properties are shown in figure \ref{fig:preds_vs_true}. We observe that the network predictions are biased towards the mean of the distribution. This is commonly observed in neural network predictions, as the networks are trained to minimize a mean error over multiple examples. The plots show that predictions for values corresponding to CCs (low values of 
log $t_{cool, 0}$, high values of log $n_{e,0}$, low values of log $K_0$, high values of  log $C_{SB}$ and high values of log $\alpha$) are further away from the true values and contain more scatter, which is in part due to the imbalance in the dataset.\\

For the cooling time (top left panel of figure \ref{fig:preds_vs_true}),  we have a good correspondence between the predictions and true values on the training set as well as the test set. We compute the average of the percentage errors between the predictions and the true values over all the examples in the training dataset as well as the test dataset. We obtain a mean percentage error of 0.7\% on the training set, and 1.8\% on the test set. \\

For the density (second panel of the first row of figure \ref{fig:preds_vs_true}), we have a good correpondance up to about log $n_{e,0}$=-2.0, and the network has trouble predicting higher values. We obtain a mean percentage error of 3.9\% on the training set and 7.7\% on the test set.\\

For the entropy (top right panel of figure \ref{fig:preds_vs_true}), we have a larger scatter. The correpondance is good down to about log $K_0$=1.3, and the network struggles predicting lower values. We obtain a mean percentage error of 4.9\% on the trainig set, and 11.4\% on the test set.\\

For the concentration (bottom left panel of figure \ref{fig:preds_vs_true}), the predictions are very biased towards the mean for both ends of the distribution. This might be because on a log scale, both ends of the concentration distribution have very few clusters. At the center of the distribution, we are able to achieve a good correlation and a relatively small scatter for the test set, when comparing to other metrics. We obtain a mean percentage error of 6.8\% on the training set and 8.0\% on the test set.\\

For the cuspiness predictions (second panel in the second row of of figure \ref{fig:preds_vs_true}), we are not able to achieve good performance on the test set. The correlation is good for the training set, but the scatter on the test set is very large, and there doesn't seem to be a clear correlation between predictions and true values for the red data points. We obtain a mean percentage error of 14.3\% on the training set and 96.6\% on the test set. This is a clear case of overfitting; better performance might be achieved through stricter regularization techniques.\\

\begin{figure}
    \centering
    \vspace{0.5cm}
    \includegraphics[width=0.5\textwidth]{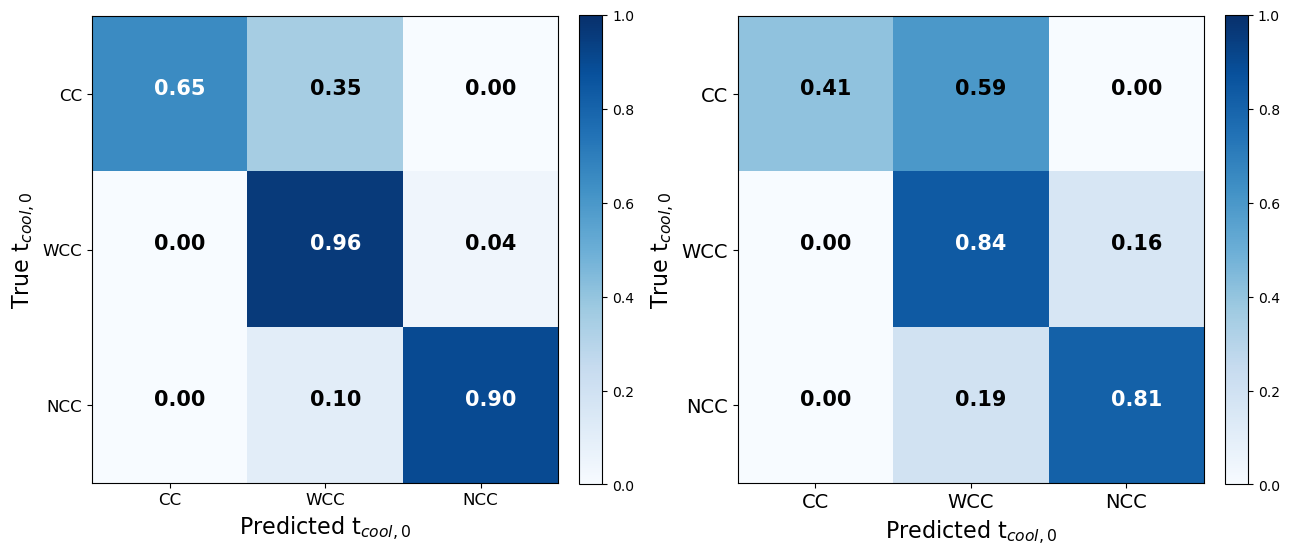}
    \caption{Normalized confusion matrices for the central cooling time predictions, for the training set (left) as well as the test set (right). The matrices are normalized along the true values. These matrices allow us to visualize the performance of the model for each metric. 
    We can see the true positives, true negatives, false positives and false negatives for each of the categories. We present matrices for both the training set and the test set to compare performances.}
    \label{fig:tcool_matrix}
\end{figure}

\begin{figure}
    \centering
    \vspace{0.5cm}
    \includegraphics[width=0.5\textwidth]{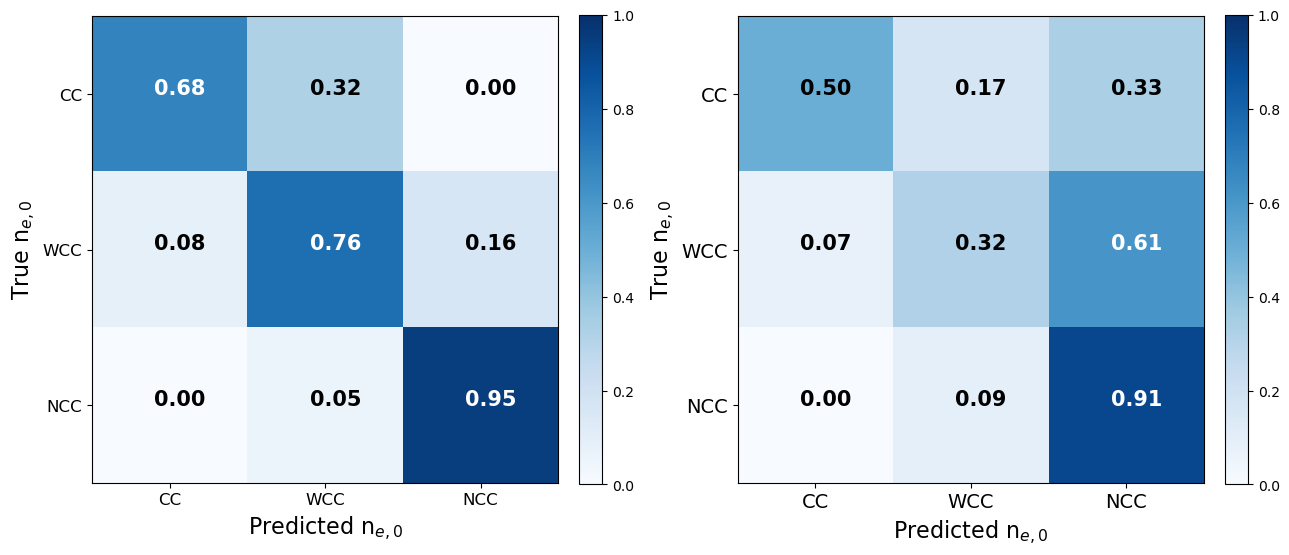}
    \caption{Normalized confusion matrices for the central electron density predictions, for the training set (left) as well as the test set (right). The matrices are normalized along the true values.}
    \label{fig:density_matrix}
\end{figure}

\begin{figure}
    \centering
    \includegraphics[width=0.5\textwidth]{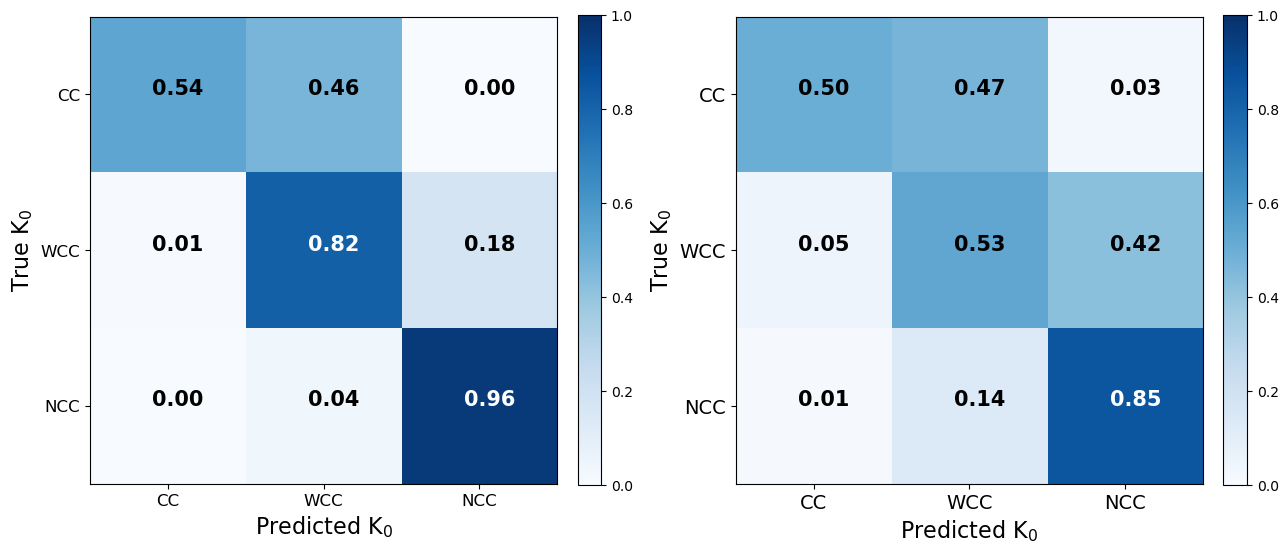}
    \caption{Normalized confusion matrices for the central entropy excess, for the training set (left) as well as the test set (right). The matrices are normalized along the true values.}
    \label{fig:entropy_matrix}
\end{figure}

\begin{figure}
    \centering
    \includegraphics[width=0.5\textwidth]{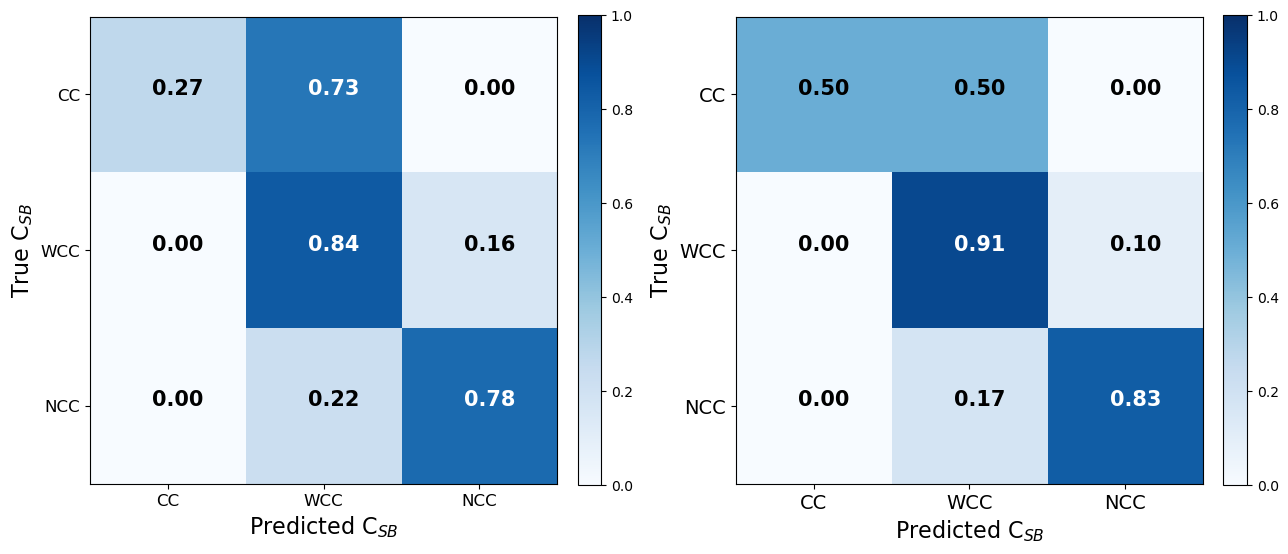}
    \caption{Normalized confusion matrices for the concentration predictions, for the training set (left) as well as the test set (right). The matrices are normalized along the true values.}
    \label{fig:conc_matrix}
\end{figure}

\begin{figure}
    \centering
    \includegraphics[width=0.5\textwidth]{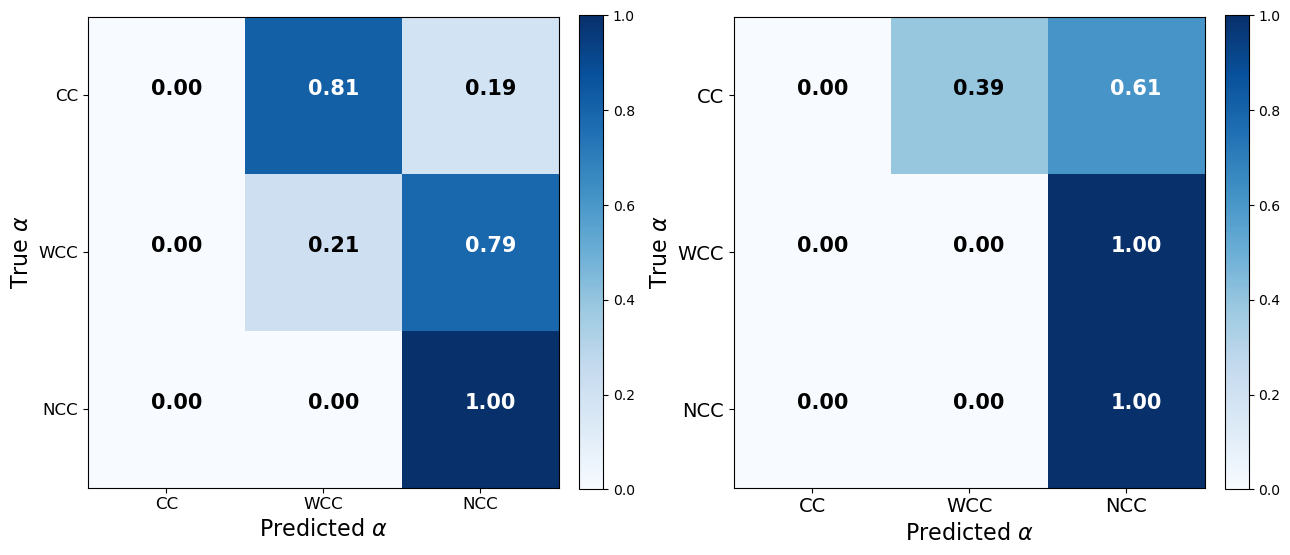}
    \caption{Normalized confusion matrices for the cuspiness predictions, for the training set (left) as well as the test set (right). The matrices are normalized along the true values.}
    \label{fig:cusp_matrix}
\end{figure}

From the results in figure \ref{fig:preds_vs_true}, we can convert the continuous predictions into class predictions, based on the cutoffs for CC/WCC/NCC clusters for each metric. We can then construct confusion matrices to visualize the performance of this classification. The confusion matrices for each metric are presented in figures \ref{fig:tcool_matrix} through \ref{fig:cusp_matrix}.\\

For every matrix, we can calculate the balanced accuracy (BAcc) for each class, defined as:

\begin{equation}
    \text{BAcc} = \frac{1}{2}\left(\frac{tp}{tp+fn} + \frac{tn}{tn + fp}\right)
\vspace{5mm}
\end{equation}
where $tp$, $tn$, $fp$ and $fn$ are true positives, true negatives, false positives and false negatives. These numbers are defined as numbers of samples, and are not normalized in the calculation of the BAcc. The balanced accuracy for each metric is shown in table \ref{tab:bacc}. Based on these results, the best-performing metrics are the central cooling time and the concentration.\\

We can compare the cooling time balanced accuracy with the results from S20. They tackle a classification problem, predicting the galaxy cluster category (CC/WCC/NCC) for mock Chandra images obtained from the $z=0$ snapshot of the TNG300 simulations. Their sample contains 318 clusters, and they use the 3D central cooling time to create class labels. They get a balanced accuracy of 0.92 for CCs, 0.81 for WCCs and 0.83 for NCCs. 10\% of clusters in their dataset are CCs, 61\% are WCCs and 29\% are NCCS. Their dataset is therefore more balanced than ours, which tends to improve performance. Finally, their network architecture is similar to ours, except for the output layers; their network predicts a class whereas our network predicts simultaneously the numerical values of the five classification metrics. Their network is also pre-trained for a classification task on the ImageNet dataset. We discuss this further in Section \ref{sec:discussion}.\\

\begin{table}[h]
\hspace{-5em}
\begin{tabular}{@{}|l|l|l|l|l|} 

\hline

\textbf{}                                 & \textbf{CC} & \textbf{WCC} & \textbf{NCC} & \textbf{Average} \\ \hline
Central cooling time $t_{\text{cool}, 0}$ & 0.70        & 0.81         & 0.83         & 0.78             \\ \hline
Central electron density $n_{e,0}$        & 0.75        & 0.61         & 0.68         & 0.68             \\ \hline
Central entropy excess $K_0$              & 0.74        & 0.67         & 0.80         & 0.74             \\ \hline
Concentration $C_{SB}$                    & 0.75        & 0.86         & 0.87         & 0.83            \\ \hline
Cuspiness $\alpha$                        & 0.50        & 0.49         & 0.58         & 0.52             \\ \hline
\end{tabular}\\

\caption{Balanced accuracy (BAcc) calculated on the test set for each class (CC, WCC, NCC) as well as the average, for each classification metric. The BAcc is an accuracy measure defined between 0 and 1, with 1 being the most accurate.}
\label{tab:bacc}
\end{table}

\vspace{10mm}
\section{Discussion}
\label{sec:discussion}

In this paper, we have used galaxy cluster data from the IllustrisTNG simulations to produce mock X-ray images as well as to calculate five classification metrics. We then train a ResNet model to predict those five properties from a downsampled X-ray image of the galaxy cluster. 

\subsection{Simulation-based inference}
\label{sec:sbi}

Once the network is trained, we can use it to get predictions on new data. Furthermore, we can obtain a probability distribution over the possible values using the formalism of simulation-based inference (SBI, e.g. \citealt{Cranmer_2020}), also referred to as likelihood-free inference (LFI). This techniques allows us to obtain a probability distribution over possible values, rather than just a single point prediction. In this context, we have a model with parameters (latent variables) $\theta$, generating data $x$. We would like to perform inference on $\theta$ given observations $\{x\}_i$ to get $p(\theta|\{x\}_i)$. This can usually be done with Bayes' theorem:

\begin{equation}
p(\theta|x) \propto \frac{p(x|\theta)p(\theta)}{p(x)},
\vspace{5mm}
\end{equation}
where $x$ is the observed data,  $p(\theta|x)$ is the posterior distribution of the model parameters given $x$, $p(x|\theta)$ is the likelihood of the data, $p(\theta)$ is the prior on the model parameters. The likelihood $p(x|\theta)$, however, is often unknown. 
SBI is useful in those cases, as it does not require an explicit expression for the likelihood. Instead, it only relies on having a model to generate $\{\theta,x\}$ pairs, that are used to fit a continuous function $f = \hat{p} (\theta|x)$ that approximates the distribution $p(\theta|x)$ for those pairs. This is done by first using the dataset $\{\theta,x\}_i$ as training data to model a joint distribution of the parameters and the data $p(x, \theta)$. Then, the posterior for new data examples can be obtained by evaluating the joint distribution at the observed data, yielding  $\hat{p} (\theta|x)$. The function $f$ can be represented as a neural network, training on $\{\theta,x\}_i$ to maximize $\Sigma_i \text{ log } p(\theta|x)$. One can use a gaussian model or a normalizing flow to ensure that $f$ is a good fit for a probability distribution, i.e. that it is smooth and normalized. Once $f$ is trained, we can sample from $\hat{p}(\theta|x)$ to build a posterior distribution for a given $x$.\\

When dealing with a regression network that predicts point estimates, SBI can be used to obtain posterior distributions for the network predictions (e.g. \citealt{legin_2023}). In our case, we wish to obtain a probability distribution over possible values of $y_{\text{true}}$, for a given value of $y_{\text{pred}}$. We use the \texttt{sbi} python package (\citealt{sbi_ref}) to perform the analysis, where the latent variable $\theta$ is $y_{\text{true}}$, $x$ is $y_{\text{pred}}$, and the learned posterior $\hat{p}(\theta|x)$ is given by $\hat{p}(y_{\text{true}}|y_{\text{pred}})$. We use the entire training dataset (all five folds) as observations for training the function $f = \hat{p}(y_{\text{true}}|y_{\text{pred}})$. We train five separate functions for the five cluster properties. \\

\begin{figure}
    \centering
    \includegraphics[width=0.5\textwidth]{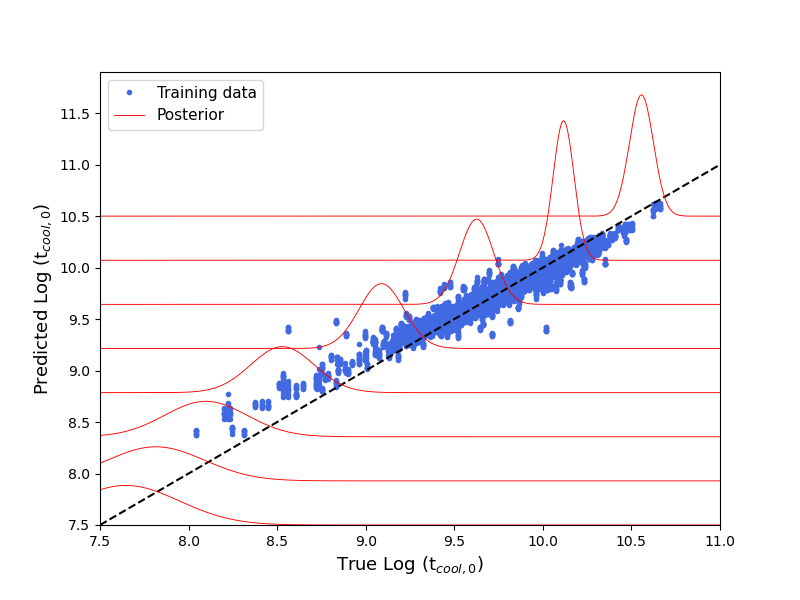}
    \caption{SBI results for the central cooling time predictions. The network predictions for the training data are plotted in blue as a function of the true labels. The black dashed line represents the true values. In red, we overplot the posterior distributions for a few predicted values of the central cooling time. We notice that the posteriors exhibit higher precision where more training data was available, as expected. }
    \label{fig:tcool_posteriors}
\end{figure}

In figure \ref{fig:tcool_posteriors}, we plot a few examples of posterior distributions (in red) for $y = \log (t_{\text{cool, 0}})$. We observe that the posterior probability distributions exhibit higher precision at the values where there is more training data and where the predictions are closer to the ground truth values, as expected. On the lower end of the distribution, the posteriors are more flat (i.e. the predictions have a higher uncertainty) as the training data points in this region have larger errors, when comparing the predictions to the true values.\\

\begin{figure*}
    \centering
    \includegraphics[width=\textwidth]{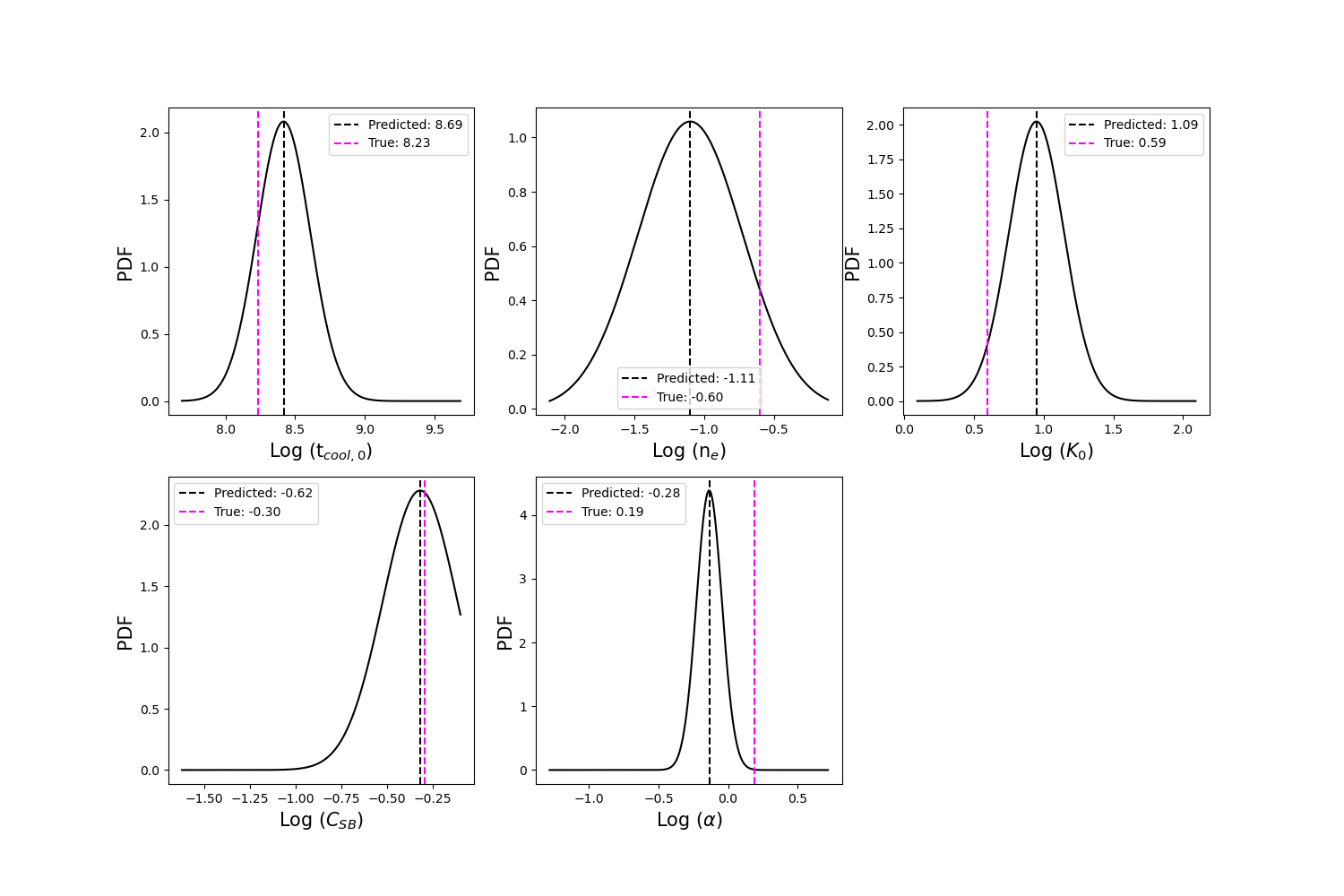}
    \caption{Example of posteriors for the five metrics, for a CC cluster. The example cluster is drawn from the test set. In each plot, the pink dashed line represents the true label and the black dashed line represents the network prediction. The solid black line represents the posterior, i.e. the probability distribution over true values for the given predicted value.}
    \label{fig:sbi_cc_exapmle}
\end{figure*}

\begin{figure*}
    \centering
    \includegraphics[width=\textwidth]{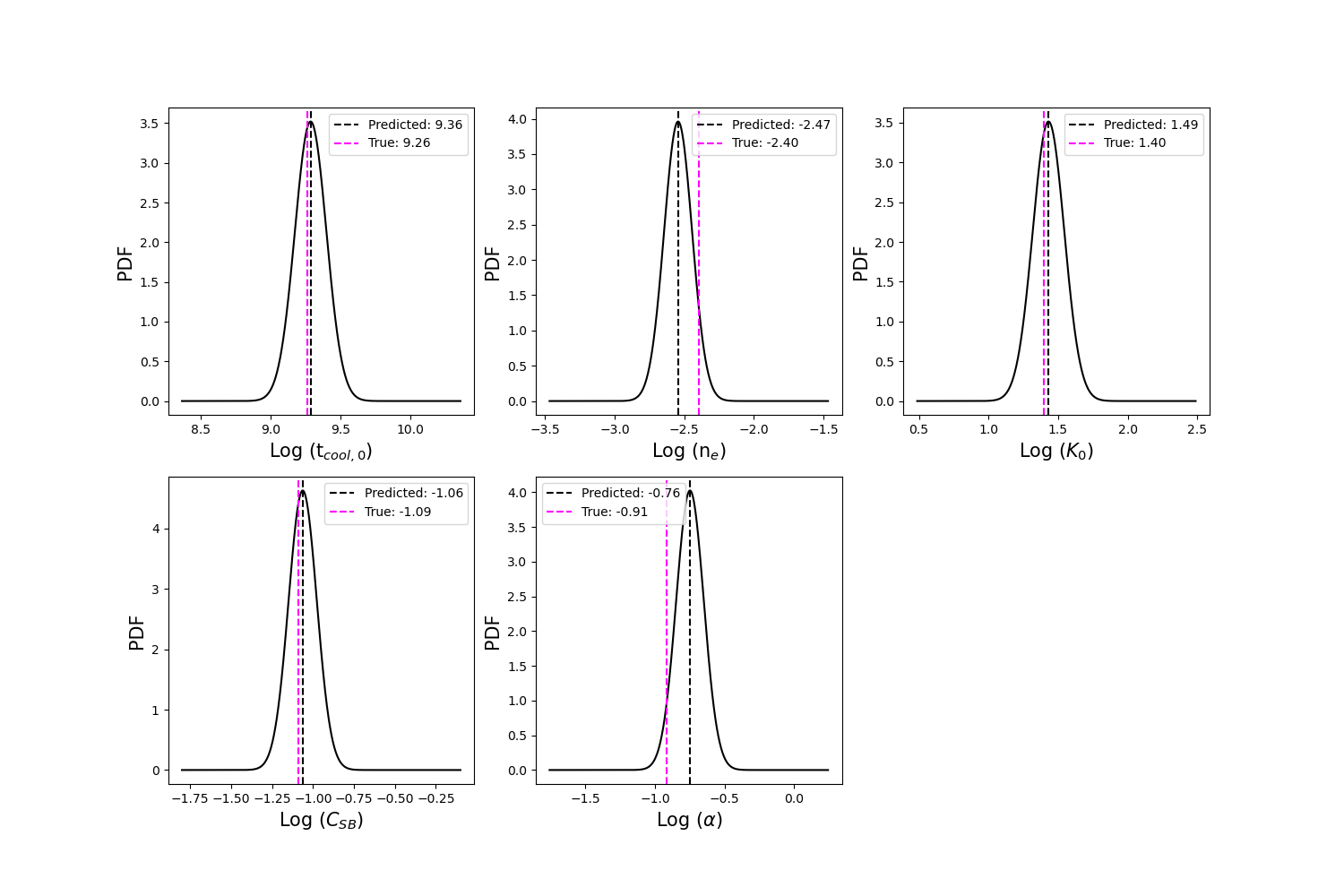}
    \caption{Same as \ref{fig:sbi_cc_exapmle} but for a NCC cluster. This example cluster is also drawn from the test set. When comparing with figure \ref{fig:sbi_cc_exapmle}, we notice that the posteriors are more narrow for the NCC cluster, which indicates that the network is more confident in the prediction. We also notice that the true values (in pink) are closer to the peak of the probability distribution.}
    \label{fig:sbi_ncc_example}
\end{figure*}

Once the posterior distribution is calculated, we can examine the probability distributions for all five metrics for the same example. This is done in figure \ref{fig:sbi_cc_exapmle} for a cool core cluster and in figure \ref{fig:sbi_ncc_example} for a non-cool core cluster. The clusters used in these figures are drawn from the test set. When comparing figures \ref{fig:sbi_cc_exapmle} and \ref{fig:sbi_ncc_example}, we clearly see that the posterior probability distributions have higher precision for the NCC example. This is expected, as there is more data for NCC clusters and therefore more information to build more constraining posterior distributions. We also see that for the NCC example, the true values generally follow the peak of the posterior distribution. For the 
density and the cuspiness, the true values have slightly lower probability. For the CC cluster, the posterior distributions are less precise and the true values (in pink) often have low probability. The probability distribution for a single example can also be used to determine what are the chances of that observation belonging to a cluster category (CC/WCC/NCC). The total area under the curve sums to one, and therefore the area under the curve that is within the range for the specified category on the x-axis indicates the probability that the cluster belongs to that category.\\

\subsection{Fraction of cool cores in IllustrisTNG}

We now proceed to compare our CC fractions to other works that have been based on the IllustrisTNG simulations, as well as on observational data. Our entire sample contains 606 clusters (1818 observations) with $M_{500c}$ $>$ 10$^{13.57} M_{\odot}$. Also drawn from IllustrisTNG, the sample in S20 consists of 318 clusters with $M_{500c}$ $>$ 10$^{13.75} M_{\odot}$, whereas the sample in B18 contains 370 clusters with $M_{500c}$ $>$ 10$^{13.75} M_{\odot}$. Since we are using a different mass cutoff, we present the CC fractions in our entire sample as well as in the subset of 363 clusters with $M_{500c}$ $>$ 10$^{13.75} M_{\odot}$ for a closer comparison.\\

When using the central cooling time as the metric, we obtain a CC fraction of 3\% for the entire sample, and 2\% for the higher mass subset. S20 obtain a CC fraction of 10\%, while B18 obtain a CC fraction of 12\% with the same definition ($t_{\text{cool},0} <$ 1 Gyr). We compare with observations of the HIghest X-ray FLUx Galaxy Cluster Sample (HIFLUGCS) from the ROSAT survey, where 44\% of the clusters are CCs (\citealt{Mittal_2009}). Hence, when using the central cooling time as the metric, the CC fraction reported in our work is lower than in S20 and B18, as well as in observational data.\\

When using the central electron density as the metric, we obtain a CC fraction of 4\% for the entire sample, and 3\% for the higher mass subset. B18 report a CC fraction of 14\% with this metric. We compare with \textit{Chandra} observations from the \textit{Planck} Early Sunyaev-Zel'dovich (ESZ) sample (\citealt{Andrade_Santos_2017}) and a flux X-ray limited sample (\citealt{Andrade_Santos_2017}). Selecting clusters with z $<$ 0.25, the SZ sample has a CC fraction of 52\% and the X-ray sample has a CC fraction of 37\%. Hence, when using the central electron density as the metric, the CC fraction reported in our work is also lower than in B18 as well as in observational data.\\

When using the central entropy excess as the metric, we obtain a CC fraction of 10\% for the entire sample, and 7\% for the higher mass subset. B18 report a CC fraction of 18\% with this metric. We compare with a sample of 239 clusters taken from the \textit{Chandra} archive analyzed in \citet{cavagnolo_2009_article}, which yields a CC fraction of 46\% for the same definition ($K_0$ $<$ 30 keV cm$^2$). Therefore, when using the central entropy excess as the metric, the CC fraction reported in our work is lower than in B18 as well as in observational data.\\

When using the concentration parameter as the metric, we obtain a CC fraction of 3\% for the entire sample, and 5\% for the higher mass subset. B18 report a CC fraction of 1\% with this metric. We compare with the SZ and X-ray samples from \citet{Andrade_Santos_2017}, selecting one again clusters with z $<$ 0.25. The SZ sample has a CC fraction of 34\% and the X-ray sample has a CC fraction of 12\%. Thus, when using the concentration as the metric, the CC fraction reported in our work is higher than in B18 but lower than in observational data.\\

When using the cuspiness parameter as the metric, we obtain a CC fraction of 1\% for the entire sample as well as for the higher mass subset. B18 obtain a CC fraction of 21\% with this metric. Selecting the clusters with z $<$ 0.25 from \citet{Andrade_Santos_2017}, the SZ sample has a CC fraction of 47\% and the X-ray sample has a CC fraction of 28\%. Therefore, when using the cuspiness as the metric, the CC fraction reported in our work is lower than in B18 as well as in observational data.\\

Overall, the CC fractions in this work are lower than the ones reported in S20 and B18. While our study, as well as B18 and S20, uses the z=0 snapshot of the TNG300 simulation, there are significant differences in the methods, which could explain the discrepancy. First, the methods used to locate the center are different; B18 use the cluster's potential minimum, while we use an iterative gaussian filter on the 2D observations to find the X-ray peak, in an effirt to mimic the methods used with observational data. It is not detailed in S20 how the center is located. Also, B18 and S20 both define the central cooling time as the average within a 3D region with $r<0.012R_{500}$, while we examine the 2D cooling time profiles and perform a fit for $t_{\text{cool},0}$. We also perform a fit for the central electron density based on the 2D profiles, while B18 use the average value within $r<0.012R_{500}$. For the entropy, our method is the same as in B18; the only difference being them using 3D entropy profiles while we're working with 2D profiles. For the concentration, the X-ray luminosity is not computed the same way in our work and in B18. We use mock \textit{Chandra} images produced with \texttt{pyXSIM} and \texttt{SOXS}, whereas B18 first produce mock X-ray spectra for each gas cell and then sum the spectra with an aperture (40 kpc or 400 kpc for $C_{SB}$) to get the X-ray luminosities. Finally, the cuspiness is calculated slightly differently; B18 get the cuspiness from 3D density profiles, while we are once again using 2D profiles. All of these differences in our analysis lead to different values for the gas properties, and therefore different CC fractions.\\

In general, the CC fractions in the z=0 snapshot of the TNG300 simulations are significantly lower than in observational data. This might be due to the fact that realistic AGN feedback processes are needed to produce the observed CC fractions, but are challenging to simulate (\citealt{Morganti_2017}). The physical processes happen at very different scales, and large cluster simulations do not have high enough resolution to accurately model the small scale physics. Therefore, simplified models are employed on larger resolved scales to measure black hole accretion rates and inject the corresponding feedback energy into the surrounding gas (\citealt{Weinberger_2018_BHfeedback}). Also, AGN feedback exists in two modes, depending on the value of the accretion rate: radio mode at a low accretion rate, and quasar mode at a high accretion rate. The physics transition between the two is difficult to implement in simulations as it is poorly understood (e.g. \citealt{Weinberger_2017}). In radio mode feedback, we expect the feedback to occur in the form of jets which inflate cavities and this is very difficult to simulate in such large cosmological volumes. In TNG300, the authors instead mimic radio jets by injecting kinetic energy through a unidirectional momentum kick, which then inflates X-ray cavities similar to those seen in observations \citep{Prunier2024}.


\subsection{Metric comparison \label{sec:metric_comp}}

Based on our analysis, we conclude that the metrics most suited for machine learning analysis of X-ray images of galaxy clusters are the central cooling time and the concentration. The cooling time performs the best out of all the metrics when it comes to the regression task. It has the smallest percentage error on the test set (1.8\%) as well as the training set (0.7\%), and the second highest balanced accuracy (0.78). The concentration has the highest balanced accuracy (0.83) out of all the metrics, and it has the strongest correlation with the group index obtained with the clustering algorithm. Since the concentration is the only metric calculated directly from the X-ray image, we expect both models (the clustering algorithm and the ResNet model for regression) to perform fairly well when using the concentration as the metric. It is therefore surprising that the ResNet has trouble predicting the $C_{SB}$ values. One reason for this might be that the information needed to predict $t_{cool,0}$, $n_{e,0}$, $K_0$ and $\alpha$ is significantly different from the information needed to predict the concentration. This would lead the convolutional layers to extract features that are not well-suited to accurately predict the concentration. This can be mitigated by training a separate network for the concentration or by adding more complexity to the existing network. Another path worth exploring would be to replace the input image with its PCA representation, as done for the k-means clustering.\\

\begin{figure}[]
  \centering
{\includegraphics[width=0.5\textwidth]{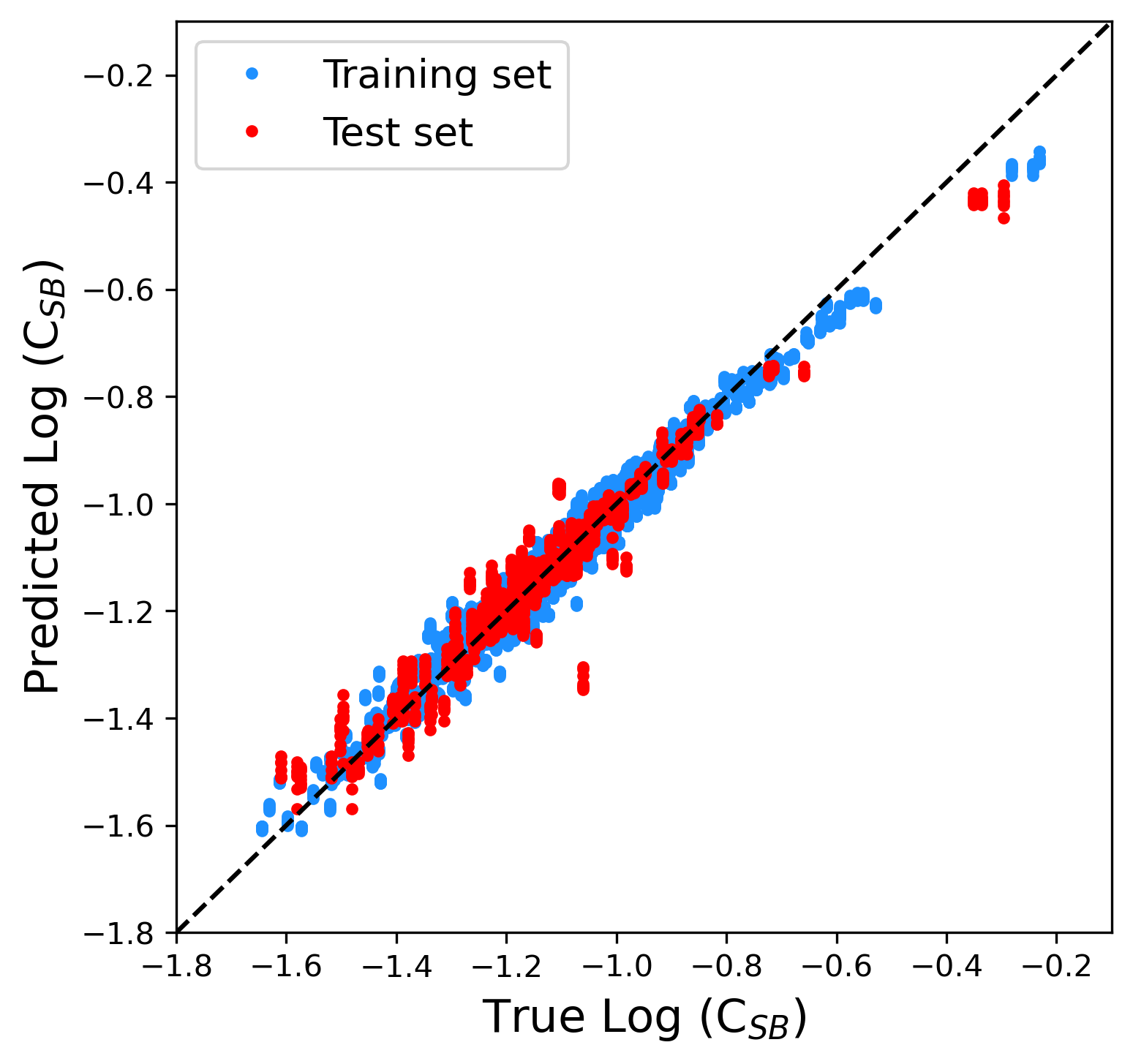}}
  \caption{Predicted values as a function of the true values for the concentration obtained with the retrained network, as described in Section \ref{sec:metric_comp}. The performance is significantly improved by retraining the network. }
  \label{fig:new_conc_pred}
\end{figure}

\begin{figure}[]
  \centering
  {\includegraphics[width=0.5\textwidth]{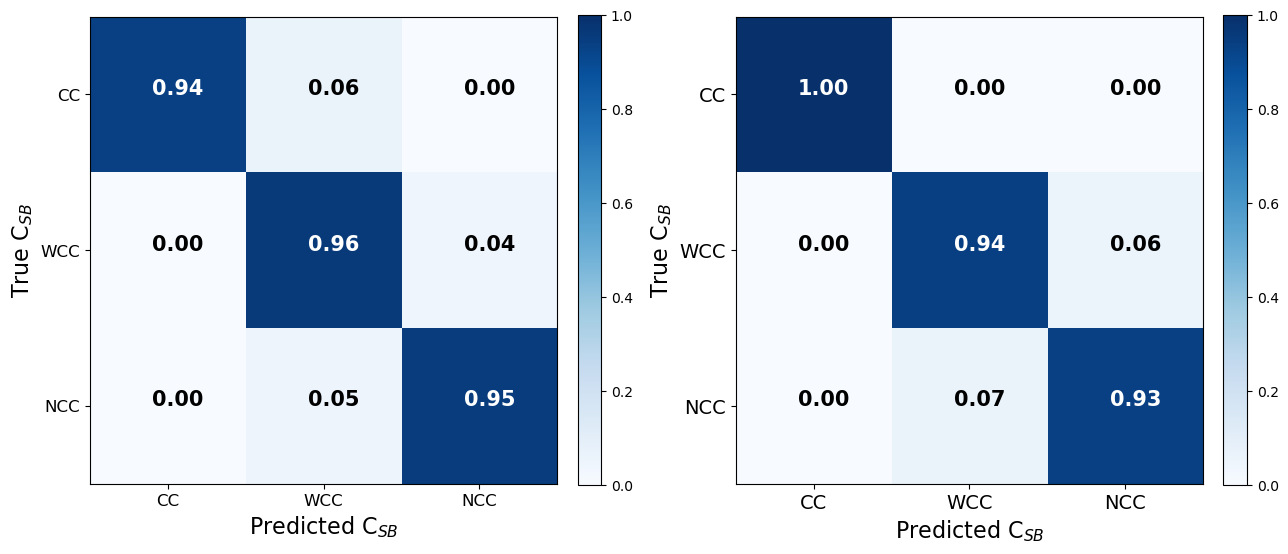}}
  \caption{Normalized confusion matrices for the concentraion for the training set (left) and the test set (right) obtained with the retrained network, as described in Section \ref{sec:metric_comp}. Once again, we notice that retraining the network to predict this metric by itself significantly improves performance. }
  \label{fig:new_conc_matrices}
\end{figure}

In an effort to improve performance for the concentration, we re-train the network only predicting the concentration values. The network is the same as the one presented in figure \ref{fig:my_resnet_model}, with the 5 neurons in the output layer being replaced by only one. The results are presented in figures \ref{fig:new_conc_pred} and \ref{fig:new_conc_matrices}. The performance is significantly improved; the correspondence between predicted and true values in the figure \ref{fig:new_conc_pred} is drastically better than the one in figure \ref{fig:preds_vs_true}. The same is true for the values in the confusion matrices in \textbf{b)} of figure \ref{fig:new_conc_matrices}: they represent a strong improvement compared to figure \ref{fig:conc_matrix}. With the retrained network, we obtain a mean percentage error of 1.6\% on the training set, and 2.9\% on the test set. We also obtain a balanced accuracy of 1.00 for CCs, 0.94 for WCCs and 0.94 for NCCs resulting in an average of 0.96.\\

For the same reason, we also re-train a separate network for the four other properties (the central cooling time, the central electron density, the central entropy excess and the cuspiness) individually. This does not result in significant change (up to 2.8\%) for the mean percentage error for these four properties. The balanced accuracy is not greatly affected (up to 0.03) for the central cooling time, the central electron density, the central entropy excess or the cuspiness. The most significant change remains to the concentration, and with the re-trained network the concentration is the metric that is most well-suited for a machine learning analysis when working with X-ray images.\\

On the other hand, the metric that performs the worst is the cuspiness $\alpha$. Visually, the predicted values on the test set for this metric do not show any correlation with the true values. It has the highest percentage error on both the trainnig set (14.3\%) and the test set (96.6\%), and it has the lowest balanced accuracy (0.52). The cuspiness is calculated from the fit obtained on the electron density profile; it is the only metric that is derived in such a secondary way. Also, the cuspiness is a very local measure of the density profile; it is calculated at exactly 0.04$R_{500}$. Moreover, when we look at the distribution of the cuspiness values, we only have 21 cool cores out of 1818 observations, making the cuspiness the least balanced metric. The significant imbalance in the dataset, as well as the extremely local calculation for this metric, could explain the poor performance.\\

Another factor to consider is the choice of $\beta$-model for the electron density profile (described in Section \ref{sec:central_ne}), which could affect our results for both the density metric as well as the cuspiness parameter. While this traditional $\beta$-model is commonly used to fit the electron density profiles of galaxy clusters, a modified version of this model with more parameters might be better suited to account for a larger diversity of surface brightness profiles and density profiles in galaxy clusters. Our results for the density metric and the cuspiness parameter might be improved by exploring alternative models for the electron density, such as the one found in \citealt{Vikhlinin_2006_icm} or \citealt{pointecouteau_2004}:

\begin{equation}
    n_e^2 (r)=\frac{n_{e,0}^2 (r/r_c)^{-\alpha} }{(1+r^2/r_c^2)^{3\beta-\alpha/2}}.
\end{equation}
As before, $n_e(r)$ is the electron density, $r$ is the radial distance from the center, $r_c$ is the core radius, $n_{e,0}$ is the central value of the electron density, and $\beta$ is a power-law index. With the addition of $\alpha$ as a new parameter, this model could potentially improve the quality of our fit and our results.
\\

\subsection{Implications for future surveys}

There are numerous steps that can be taken in order to improve the performance of the model and make it applicable to observational data. As discussed earlier, the network performance can be improved by training on a larger dataset, by adding complexity to the network, or by making the input more informative with the addition of spectral X-ray data. A larger training dataset could be sourced from two upcoming projects: TNG-Cluster and MilleniumTNG. TNG-Cluster is a spin-off series of simulations by the IllustrisTNG team, specifically designed for the study of massive galaxy clusters. It is expected to have a simulation box size of 1000 Mpc, compared to the current largest simulation box size of 300 Mpc. Millenium-TNG aims to provide better representation of dark matter structures, while having a simulation box size of 500 Mpc. Both of these simulations are expected to be made publically available in 2024.\\

A promising path to make the model applicable to observational data is to use transfer learning, which consists of using a pre-trained model on a new problem. In our case, we can use the model trained on simulated data as a starting point and fine-tune it by retraining it on a small dataset of observational data. This has been successfully done in astrophysics (e.g. \citealt{ghosh_2022}), and could greatly improve the performance on the entire dataset of observational data. \\

A large dataset of observational X-ray data will become available in the next few years, as the X-ray instrument eROSITA is expected to detect at least 100 000 massive galaxy clusters (\citealt{Merloni_2012_erosita}, \citealt{Pillepich_2012_erosita}). A first data set was released in January 2024, with more data scheduled to become available in the next few years. To analyze these large datasets, galaxy cluster experts will need assistance from machine learning tools, such as the model presented in this work. Our model is a fast an efficient way to characterize the ICM in galaxy clusters. While the model performance is not on-par with expert classification, we have produced a useful tool that provides uncertainty mesures. This is a viable path to assist experts in their analysis by isolating the samples that show large uncertainties, helping them target the more complex galaxy clusters and allowing us to make the most out of the upcoming data.\\

\section{Conclusion} \label{sec:conclu}
We have presented a deep learning approach for galaxy cluster characterization, predicting five cluster properties from mock X-ray images. The training dataset is built from 606 clusters with $M_{500c} > 10^{13.57} M_{\odot}$, resulting in 1818 observations and augmented to 14 544 images. The network is trained to learn cluster properties from a downsampled 256 $\times$ 256 image. Our method shows a mean percentage error of 1.8\% for the central cooling time, and an average balanced accuracy of 0.83 for the concentration. We have performed an SBI analysis to obtain posterior probability distributions for the network predictions. This analysis shows that the predictions for non-cool core clusters have lower uncertainty as well as lower error. Based on the network predictions as well as the probability analysis, we conclude that the cooling time and the concentration are the metrics that lend themselves the best to machine learning characterization. We also note that the concentration parameter correlates very strongly with the groups resulting from the unsupervised clustering algorithm. This indicates that the classification based on the concentration corresponds to a somewhat natural separation of galaxy cluster X-ray images. \\

Overall, we have produced a fast and statistically sound model that can be used to accelerate the analysis of large datasets. An important limiting factor in our work is the size and the range of our dataset. This can be improved by pre-training our network on a separate larger dataset of astrophysical images. We could also explore a wider range of neural network models that are better able to capture the discriminative features. In a follow-up study, we would like to consider different models for the electron density profile, as mentioned in Section \ref{sec:metric_comp}. We would also like to apply and train the network on real observations, as well as include spectral information as input, since it is already available with X-ray data. We expect this to significantly improve the performance of the model, making it a powerful tool for extracting key information from galaxy cluster observations.\\

\section*{Software}

This research made use of \texttt{Python} (\citealt{python_software}), \texttt{NumPy} (\citealt{numpy_software}), \texttt{SciPy} (\citealt{scipy_software}), \texttt{AstroPy} (\citealt{astropy_software_2013}, \citeyear{astropy_software_2018}, \citeyear{astropy_software_2022}), \texttt{Matplotlib} (\citealt{matplotlib_software}), \texttt{TensorFlow} (\citealt{tensorflow_software}), \texttt{Keras} (\citealt{keras_software}), \texttt{emcee} (\citealt{emcee_paper}), \texttt{pyXSIM} (\citealt{pyXSIM_software}), \texttt{SOXS} (\citealt{soxs_software}),\texttt{yt} (\citealt{yt_software}), \texttt{CIAO} (\citealt{ciao_paper}), \texttt{sbi} (\citealt{sbi_ref}).

\vspace{5mm}
\section*{Acknowledgments}

The authors would like to thank the IllustrisTNG team. The IllustrisTNG simulations were undertaken with compute time awarded by the Gauss Centre for Supercomputing (GCS) under GCS Large-Scale Projects GCS-ILLU and GCS-DWAR on the GCS share of the supercomputer Hazel Hen at the High Performance Computing Center Stuttgart (HLRS), as well as on the machines of the Max Planck Computing and Data Facility (MPCDF) in Garching, Germany.
\vspace{3mm}\\
This research was enabled in part by support provided by Calcul Québec (calculquebec.ca) and the Digital Research Alliance of Canada (alliancecan.ca).
\vspace{3mm}\\
This research has made use of software provided by the \textit{Chandra} X-ray Center (CXC) in the application packages CIAO.
\vspace{3mm}\\
M.S. acknowledges financial support from the Fonds de recherche du Québec – Nature et technologies (FRQNT), from IVADO, from the Center for Research in Astrophysics of Quebec (CRAQ), as well as from the physics department of Université de Montréal.
\vspace{3mm}\\
M.N. acknowledges support by NASA under award No. 80NSSC22K0821.
\vspace{3mm}\\
J.Z. is funded by the Chandra X-ray Center, which is operated by the Smithsonian Astrophysical Observatory for and on behalf of NASA under contract NAS8-03060.\\
%
\vspace{5mm}





\clearpage
\section*{Appendix}

This short appendix contains figures representing the two types of residual blocks used in the neural network, as presented in Section \ref{sec:regression}.

\begin{figure}[h]
\includegraphics[width=0.3\textwidth]{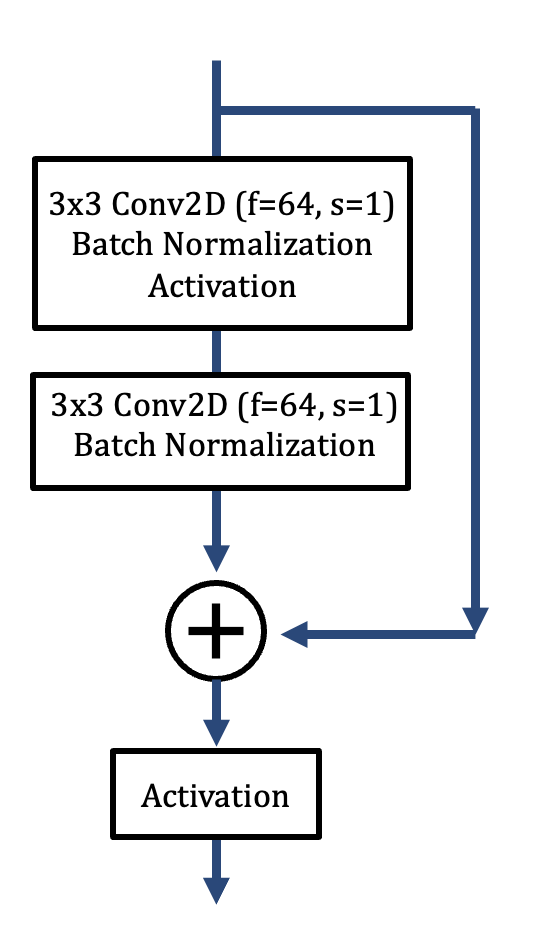}
    \caption{Residual block A. Conv2D stands for 2D convolution. $f$ and $s$ refer to the number of filters and the size of the strides.}
    \label{fig:res_block_a}
\end{figure}  
\begin{figure}[h]
\includegraphics[width=0.4\textwidth]{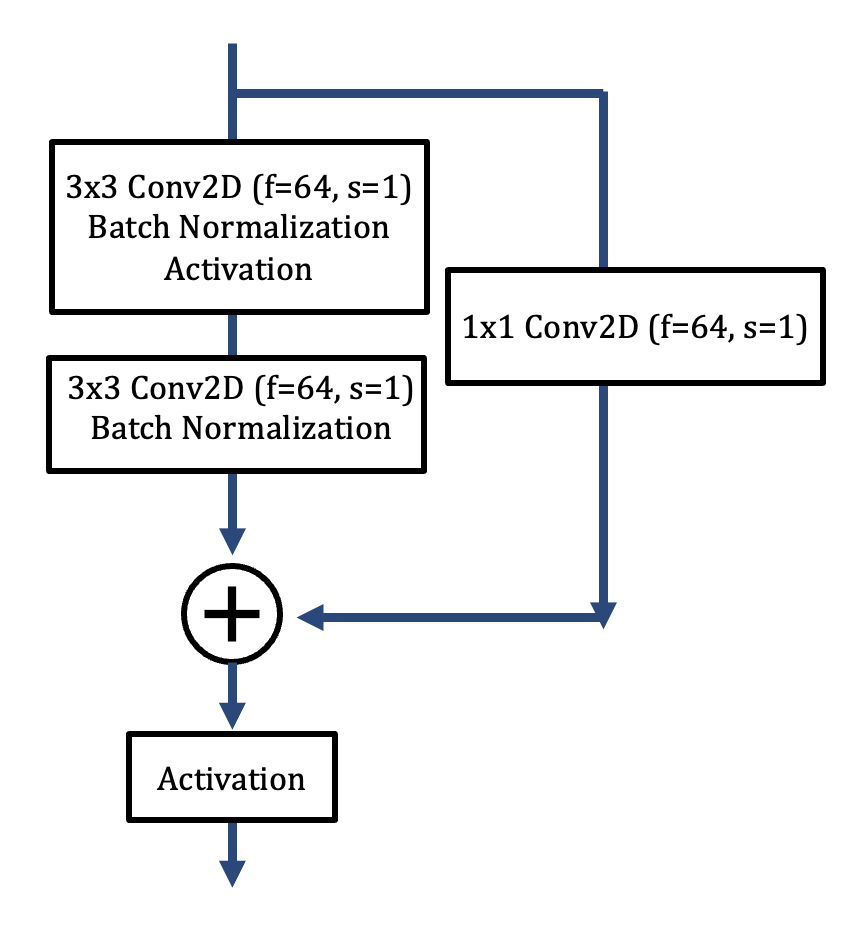}
    \caption{Residual block B. A convolutional layer is added in the skip connection.}
    \label{fig:res_block_b}
\end{figure}

\clearpage

    \label{fig:res_block_a}

\bibliography{sample631}{}
\bibliographystyle{aasjournal}



\end{document}